\documentclass[twocolumn, superscriptaddress, secnumarabic, floatfix]{revtex4}

\usepackage[version=3]{mhchem} 
\usepackage{amsmath}
\usepackage{graphicx}
\usepackage{float}
\usepackage[colorlinks=true, allcolors=blue]{hyperref}
\usepackage{rotating}
\usepackage{tablefootnote}
\usepackage{hyperref}
\usepackage{algorithmic}
\usepackage{algorithm}
\usepackage{appendix}
\usepackage{txfonts}
\usepackage{array}
\usepackage{bm}
\usepackage{amsmath}
\usepackage{color}
\usepackage{physics}
\usepackage{braket}
\usepackage{setspace}

\usepackage[caption=false]{subfig}

\def\TitleOfPaper{Exact Conditions for Ensemble Density Functional Theory}
\def\preprintlinklocation{http://dft.uci.edu + PAPER REF}

\usepackage{amsmath}
\usepackage{physics}
\usepackage{amssymb}
\usepackage{float}
\usepackage[dvipsnames]{xcolor}

\definecolor{TITLECOL}{rgb}{0.05,0.25,0.85}
\definecolor{CONTENTSCOL}{rgb}{0.1,0.2,0.7}
\definecolor{URLCOL}{rgb}{0,0.52,0.83}
\definecolor{LINKCOL}{rgb}{0.05,0.5,0}
\definecolor{CITECOL}{rgb}{0.25,0,0.48}
\definecolor{SECOL}{rgb}{0.07,0.31,0.80}
\definecolor{SSECOL}{rgb}{0.26,0.19,0.75}

\newcommand{\coloredtitle}[1]{\title{\textcolor{TITLECOL}{#1}}}
\newcommand{\coloredauthor}[1]{\author{\textcolor{CITECOL}{#1}}} 

\usepackage{graphicx}

\def\PublicationsLink{http://dft.uci.edu/publications.php}
\def\preprintlink{ \href{\preprintlinklocation}{\TitleOfPaper} }
\def\preprinttext{~}

\usepackage{fancyhdr}
\makeatletter
\fancypagestyle{titlepage}
{	\lhead{\textsc{\preprinttext\ ~ \href{\PublicationsLink}{~}}}
	\chead{}
	\rhead{\preprintlink}
	\lfoot{}}
\makeatother
\def\preprintlink{ 
	\href{\preprintlinklocation}
        {
~}
	}

\definecolor{Green}{rgb}{0.016,0.627,0}
\definecolor{Plum}{rgb}{0.17,0,0.45}
\definecolor{LBlue}{rgb}{0,0.34,0.45}
\definecolor{Sepia}{rgb}{0.37,0.17,0.02}
\definecolor{BurntOrange}{rgb}{0.78,0.39,0}

\usepackage{hyperref}
\hypersetup{
    breaklinks=true,
    bookmarksopen=true,
    bookmarksnumbered=true,
    colorlinks=true,
    linkcolor=LINKCOL,
    linktocpage=true,
    citecolor=CITECOL,
    filecolor=magenta,      
    urlcolor=URLCOL,
}

\def\bea{\begin{eqnarray}}
\def\eea{\end{eqnarray}}
\def\ben{\begin{equation}}
\def\een{\end{equation}}
\def\benu{\begin{enumerate}}
\def\enu{\end{enumerate}}

\def\bei{\begin{itemize}}
\def\eei{\end{itemize}}
\def\beit{\begin{itemize}}
\def\eit{\end{itemize}}
\def\benu{\begin{enumerate}}
\def\enu{\end{enumerate}}

\def\n{n}

\def\sss{\scriptscriptstyle\rm}

\def\w{_{w}}
\def\cw{_{{\sss C},w}}

\def\sw{_{{\sss S},w}}
\def\hxw{_{{\sss HX},w}}

\def\eew{_{{\rm ee},w}}

\def\hxcw{_{{\sss HXC},w}}
\def\g{_\gamma}

\def\wg{_{w,\gamma}}

\def\l{^\lambda}

\def\hatT{{\hat T}}
\def\hatV{{\hat V}}
\def\hatH{{\hat H}}
\def\hatG{{\hat \Gamma}}
\def\1var{(\bx_1...\bx\N)}

\def\br{{\bf r}}

\def\bx{{x}}

\def\c{_{\sss C}}

\def\N{_{\sss N}}

\def\HF{^{\rm HF}}

\def\ee{_{\rm ee}}

\def\td{time-dependent~}

\def\sph_int{ {\int d^3 r}}

\graphicspath{{figures/}}

\begin{document}
\sf
\coloredtitle{\TitleOfPaper}


\coloredauthor{Thais R. Scott}
\affiliation{Department of Chemistry, University of California, Irvine, CA 92697}

\coloredauthor{John Kozlowski}
\affiliation{Department of Chemistry, University of California, Irvine, CA 92697}

\coloredauthor{Steven Crisostomo}
\affiliation{Department of Physics \& Astronomy, University of California, Irvine, CA 92697}

\coloredauthor{Aurora  Pribram-Jones}
\affiliation{Department of Chemistry, University of California, Merced, CA 95343}

\coloredauthor{Kieron Burke}
\affiliation{Department of Chemistry, University of California, Irvine, CA 92697}
\affiliation{Department of Physics \& Astronomy, University of California, Irvine, CA 92697}

\date{\today}

\begin{abstract}
Ensemble density functional theory (EDFT) is a promising alternative to time-dependent density functional theory for computing electronic excitation energies. Using coordinate scaling, we prove several fundamental exact conditions in EDFT, and illustrate them on the exact singlet bi-ensemble of the Hubbard dimer.  Several approximations violate these conditions and some ground-state conditions from quantum chemistry do not generalize to EDFT.  The strong-correlation limit is derived for the dimer, revealing weight-dependent derivative discontinuities in EDFT.

\end{abstract}
\maketitle
\begin{figure}
    \centering
     \includegraphics[width=0.487\textwidth]{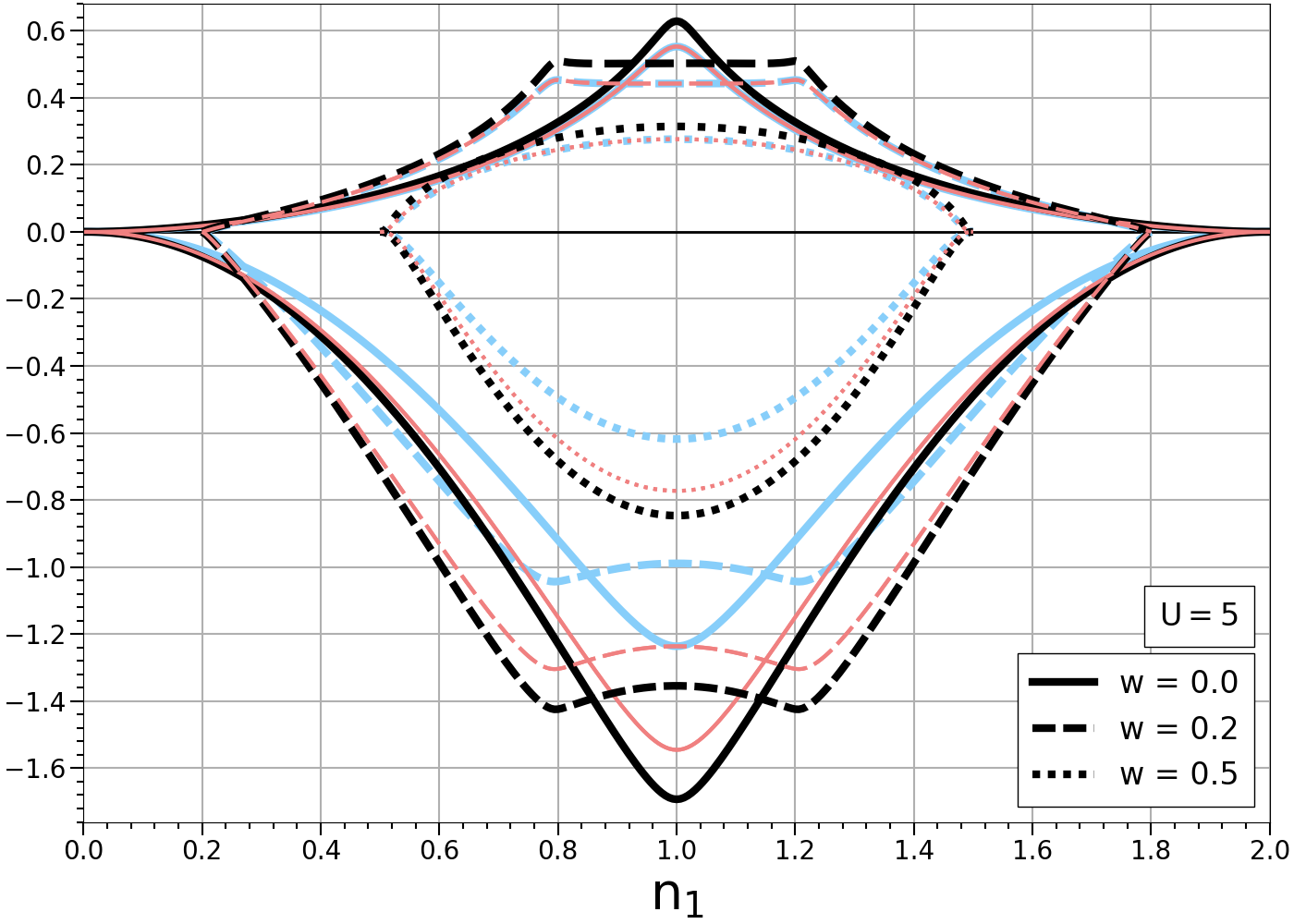}
    \caption{The Hubbard dimer singlet bi-ensemble correlation energies (negative values) and kinetic contribution (positive values) for $U = 4$ (light blue) and $U = 5$ (black) as a function of site-occupation and different weights. Red curves deduced from $U=4$ constrain the $U = 5$ curve via Eq. \ref{hubbard_uc}.}
    \label{fig-1}
\end{figure}

Sophisticated functional approximations and a relatively low computational cost have made density functional theory~\cite{hohenberg1964inhomogeneous,kohn1965self} (DFT) the prevailing method used in electronic structure calculations.~\cite{burke2012perspective, becke2014perspective, jones2015density, morgante2020devil, ullrich2011time} Currently, the most popular way to access excited states in the DFT formalism is through \td DFT (TDDFT),~\cite{runge1984density, casida1996tddft, marques2012fundamentals, ullrich2011time, maitra2016perspective} which has been used to predict electronic excitation spectra among other properties. Although TDDFT has been incredibly successful,~\cite{maitra2016perspective}  standard approximations fail to replicate charge-transfer excitation energies~\cite{tozer2003relationship}, correctly locate conical intersections~\cite{levine2006conical}  or recover double excitations~\cite{maitra2016perspective} without an \textit{ad hoc} dressing.~\cite{huix2011assessment}
\break
\\
\null \quad A less well-known but comparably rigorous alternative to TDDFT is ensemble density functional theory~\cite{theophilou1979energy, gross1988rayleigh, gross1988density} (EDFT), which is currently experiencing a renaissance. ~\cite{yang2014exact, pribram2014excitations, yang2017direct, gould2017hartree, deur2019ground, deur2017exact, gould2019density, fromager2020individual, gould2020density, gould2020ensemble, gould2021ensemble, gould2020approximately, loos2020weight, marut2020weight, gould2021double, yang2021second, cernatic2022ensemble,sagredo2018accurate} As the EDFT field is revived, it is important to find exact conditions that can be enforced on newly developed EDFT approximations.  This is especially important in EDFT, where the choice of ensemble weights is unlimited (assuming they are normalized and are monotonically non-increasing with energy) and can significantly impact the accuracy of the energies. Exact conditions have been essential in the development of accurate functionals in ground-state DFT, and we expect them to be more 
critical in EDFT.~\cite{perdew1996generalized, levy1985hellmann,pederson2023reassessing}
\break
\\
\null \quad Here, several exact conditions for EDFT are proven and illustrated. We generalize coordinate scaling inequalities and equalities of the exchange and correlation energies and the concavity condition to ensembles. Using the Hubbard dimer, we show examples of each foundational condition and examine approximations in EDFT, finding examples of compliance and violation. Fig \ref{fig-1} illustrates some of these conditions nicely.~\cite{hubbard1963electron} It shows the limits (red) one can place on the $U=5$ dimer (black) from results for $U=4$ (blue), using one of our inequalities. The rest of this paper explains the behavior of these curves, including non-monotonicity with weight and their shapes for large $U$. These exact results provide examples of the many ways in which EDFT can differ from ground-state DFT.


EDFT is a formally exact generalization of ground-state KS-DFT, where the ensemble consists of several eigenstates of an $N$-electron system. Consider any ensemble density matrix, $\hatG\w$, of the form
\ben
    \hatG\w (\br_{1}...\br_{N},\br'_{1}...\br'_{N} ) = \sum_{m=0}^{M} w_{m}\ket{\Psi_{m}(\br_{1}...\br_{N})}\bra{\Psi_{m}(\br'_{1}...\br'_{N})},
    \label{ensemble_density_matrix}
\een
where $\Psi_m$ are any orthonormal wave functions, and 
$w_{m}$ are positive monotonically non-increasing weights that are normalized. 
 The expectation value of any operator $\hat{A}$ is then
\ben
    A[\hatG\w] = \text{Tr}\{\hatG\w\hat{A}\} = \sum_{m=0}^{M} w_{m} \bra{\Psi_{m}}\hat{A}\ket{\Psi_{m}}.
    \label{expectation_value_operator_A}
\een
An ensemble energy is then the variational minimum of the Hamiltonian, yielding
\ben
    E\w = \min_{\Gamma\w} \text{Tr}\{ \hatG\w\hatH\},
    \label{ensemble_energy}
\een
where $m$ labels the eigenstates, in order of increasing energy, and $E_m$  are the eigenvalues. Transition energies can be deduced from differences between ensemble calculations of differing weights.~\cite{oliveira1988density}
EDFT tells us that there exists a $w$-dependent density functional
\ben
    F\w[n] = \min_{\Gamma\w\rightarrow \n} \text{Tr}\{\hatG\w(\hatT+\hatV \ee)\},
    \label{ensemble_functional}
\een
where $\hatT$ is the kinetic energy operator and $\hatV \ee$
is the electron-electron repulsion.  We denote the minimizer 
by $\Gamma\w[\n]$. Then 
\ben
E\w = \min_{\n}\left\{ F\w[\n]+\int \n(\br)v(\br) d\br\right\},
\label{ensemble_energy_2}
\een
where $v(\br)$ is the external potential. Any expectation value can be converted into a density functional via $A[n]=A[\Gamma\w[n]]$ . The minimizing density is
\ben
    \n\w (\br) = \sum_{m=0}^{M} w_{m} n_{m}(\br),
    \label{minimizing_density}
\een
where $\n_m(\br)$ is the density of the $m$-th level.

A key facet of EDFT is that the equivalence between the exact density and the non-interacting KS density is only true for the ensemble average, and it is not necessarily true for the individual densities within the weighted sum. The following conditions are true only for the ensemble energy, not the individual excited-state energies.

Uniform coordinate scaling has been responsible for multiple advances in DFT. However, coordinate scaling investigations in EDFT have thus far only been used to define the adiabatic connection formula for the exchange-correlation energy~\cite{nagy1995coordinate} or examining the behavior of EDFT in the low-density and high-density regimes, without formal theorems based on scaling.~\cite{gould2023electronic} Additional work on foundational theorems include the virial theorem for EDFT by Nagy \cite{nagy2013local,nagy2002theory,nagy2011functional} and the signs of correlation energy components, by Pribram-Jones {\em et al}.~\cite{pribram2014excitations} We build on this foundation by deriving uniform scaling inequalities based on the variational definition of the ensemble functional.~\cite{levy1985hellmann,pittalis2011exact} We also provide numerical verification and proofs of the basic principles and some additional exact conditions. 

 We use norm-preserving homogeneous scaling of the coordinate \br $\rightarrow \gamma \br$ with $0<\gamma<\infty$. The scaled density matrix is defined as 
\ben
    \Gamma\wg(\br_{1}...\br'_{N}) := \gamma^{3N} \,\Gamma\w
(\gamma\br_{1}...\gamma\br'_{N}),
    \label{scaled_density_matrix}
\een
and a scaled density is $\n_\gamma(\br)=\gamma^3\n(\gamma\br)$.  Trivially,
\ben
T[\Gamma\wg]=\gamma^2 T[\Gamma\w],~~~V\ee[\Gamma\wg]=\gamma \,V\ee[\Gamma\w].
\label{vee_and_t_scaling}
\een
Because these scale differently, $\Gamma\wg[\n] \neq \Gamma\w[\n_\gamma]$.
By the variational principle, $F[n\wg] \leq  F[\hatG\wg[n]]$,
which gives the fundamental inequality of scaling,
\ben
    T\w[\n\g] +V\eew [\n\g] \leq \gamma^{2}T\w[\n] + \gamma V\eew [\n]. 
    \label{fund_ineq_scal-1}
\een
Manipulation of this formula yields, for $\gamma \geq 1$, \cite{levy1985hellmann}
\ben
    T\w[n\g] \leq \gamma^{2} T\w[n],~~~~
    V\eew[n\g]\geq \gamma V\eew[n],
    ~~~~\gamma \geq 1
    \label{separated_vee_t_inequalities}
\een
and setting $\gamma \rightarrow 1/\gamma$ yields results for $\gamma \leq 1$.

Next, we turn to the KS scheme, used in modern EDFT approaches. Here $F\w[n] = T\sw[n] + E\hxcw[n]$ where $T\sw$ is the KS kinetic energy and $E\hxcw$ is the Hartree-exchange-correlation. Because there is no interaction, 
\ben
    T\sw [n\g] = \gamma^{2} T\sw [n].
    \label{kinetic_ks_equality}
\een
Moreover, because the Hartree-exchange is linear
in the scaling parameter:
\ben
    E\hxw [n\g] = \gamma E\hxw [n].
    \label{hx_equality}
\een
In EDFT, separation of Hartree from exchange is more complicated than in ground-state DFT.
\cite{gould2023electronic,yang2021second,gould2019density} 
Subtracting these larger energies following the usual procedure from ground-state DFT \cite{levy1985hellmann} 
yields, for $\gamma \geq 1$,
\ben
    T\cw[n\g] \leq \gamma^{2} T\cw[n],~~~~
E\cw[n\g] \geq \gamma E\cw [n], 
~~~\gamma \geq 1
\label{uc-ineq}
\een

\noindent where $E\cw[\n]$ is the correlation energy, and
 $T\cw=T\w -T\sw$ is its kinetic contribution.  
Considering $\gamma=1+\epsilon$ in Eq.\ref{uc-ineq}, and taking $\epsilon \rightarrow 0$, yields differential versions of Eq. \ref{uc-ineq}:
\ben
\frac{d}{d \gamma} \left\{\frac{T\cw[n\g]}{\gamma^2}\right\}\leq 0,
~~~~
\frac{d}{d \gamma} \left\{\frac{E\cw[n\g]}{\gamma}\right\}\geq 0
\label{uc-ineq2}
\een
Combining these using Nagy's generalization (Eq.~24 of Ref.~\citenum{nagy2002theory}) of the ground-state
equality
\begin{equation}
    \left.\frac{dE\cw[n\g]}{d\gamma}\right|_{\gamma=1} = E\cw[n] + T\cw[n],
    \label{deriv_at_one}
\end{equation}
we find
\ben
\bigg( 2 - 2\gamma\frac{d}{d \gamma} + \gamma^{2}\frac{d^{2}}{d \gamma^{2}} \bigg) \, E\cw[\n\g] \leq 0,
\label{deriv_2}
\een
the condition for concavity in the ensemble correlation energy. This is the ensemble form of Eq.~40 in Ref.~\citenum{levy1993tight}. Eqs. \ref{fund_ineq_scal-1}, \ref{uc-ineq}, and \ref{deriv_2} are primary results of the current work, being the ensemble generalizations of their ground-state analogs.

An immediate application of Eq. \ref{hx_equality} is to extract the HX component from any HXC approximation.
As the conditions limit growth with $\gamma$,
\ben
    E\hxw[\n] = \lim_{\gamma \rightarrow \infty} E\hxcw [n\g]/\gamma,
    \label{high-dens-x}
\een
an exact condition which can prove useful for separating HX from C components.~\cite{gould2017hartree,gould2023electronic}

\begin{figure}
    \centering
     \includegraphics[width=0.487\textwidth]{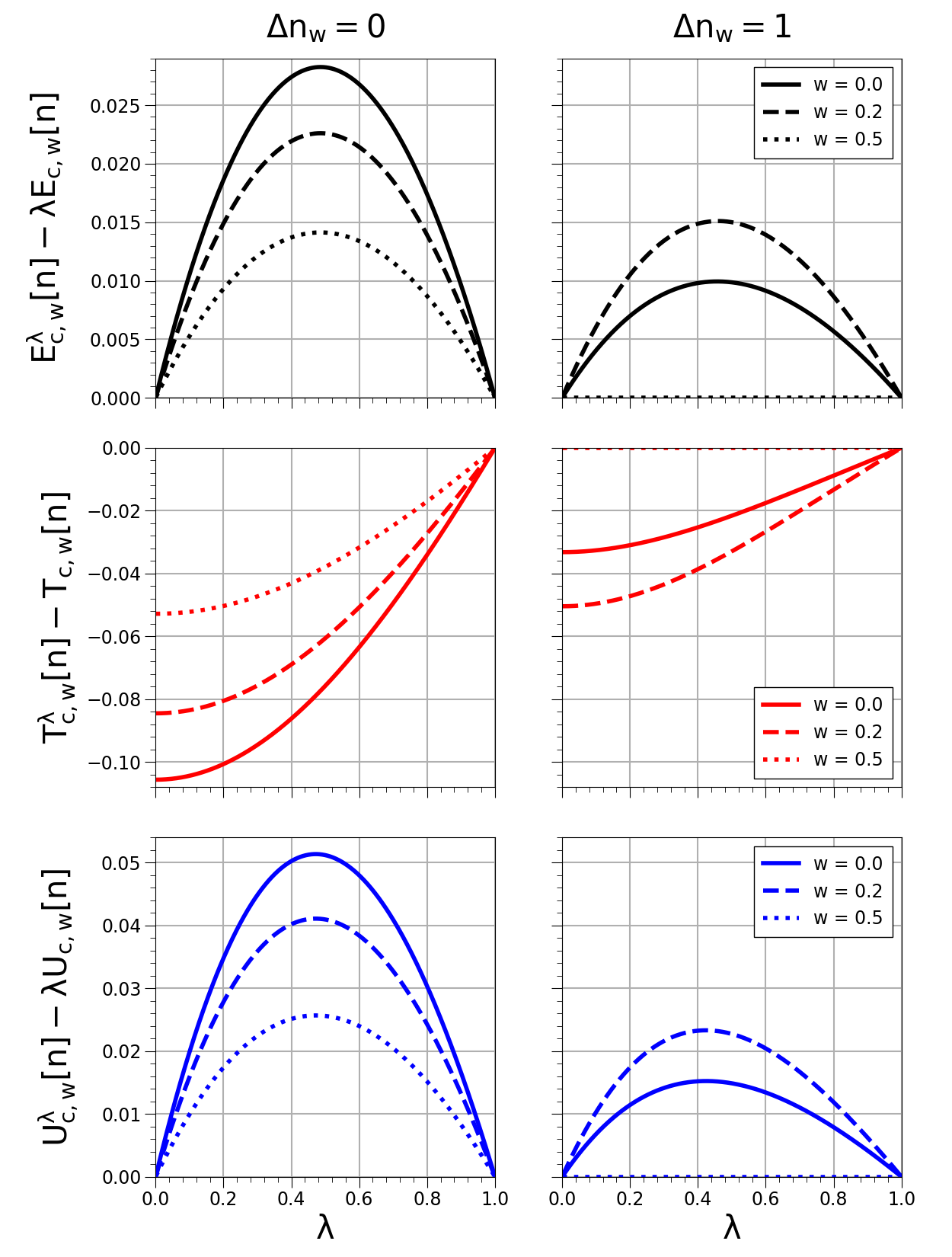}
    \caption{Correlation inequalities (Eq.~\ref{hubbard_uc}) for the total (top), kinetic (middle), and potential (bottom) correlation energies, depicted by varying $\lambda$ in the Hubbard dimer bi-ensemble with $U = 1$. More cases are provided in Figs. S6-S7 of the supplemental material.}
    \label{fig:CorrScaling}
\end{figure}

To conclude this section, we 
use the pioneering relationship between coupling constant and coordinate scaling.  Defining $\lambda$ dependence via
\ben
F\l\w [n] = \min_{\Gamma\w \rightarrow  n} \text{Tr}\{\Gamma\w(\hatT + \lambda\hatV\ee)\},
\een
Nagy showed~\cite{nagy1995coordinate}
\ben
E\hxcw\l[\n] = \lambda^2 E\hxcw[\n_{1/\lambda}].
\label{gamma_to_lambda}
\een
Using Eq.~\ref{gamma_to_lambda}, it is possible to rewrite all results given in terms of scaled densities as $\lambda$-dependent relations.
Such relations are well known and much used in ground-state DFT, via the adiabatic connection formalism. \cite{harris1984adiabatic,ernzerhof1996construction}
For real-space Hamiltonians, these relations are simply a rewriting of the scaling relations in 
a more popular form, but they also apply to 
lattice Hamiltonians, where scaling is not possible. Converting from scaling in Eq.~\ref{deriv_at_one} gives
\ben
T^{\lambda}\cw [n] = E^{\lambda}\cw [n] - \lambda \frac{d E^{\lambda}\cw[n]}{d \lambda}.
\label{eq:catherine}
\een
The scaling inequalities (Eqs.~\ref{uc-ineq}) become
\ben
     T\cw\l[\n] \leq  T\cw[\n],
     ~~~
    E\cw\l[\n] \geq \lambda E\cw[\n],
    ~~~~
    \lambda \leq 1,
    \label{hubbard_uc}
\een
with differential versions
\ben
\frac{dT\cw\l[\n]}{d\lambda} \leq 0,
~~~
E\cw\l[\n] \geq \lambda\, \frac{dE\cw\l[\n]}{d\lambda},
\label{hubbard_uc2}
\een
while Eq.~
\ref{deriv_2} becomes quite simply:
\ben
    \frac{d^{2} E\cw\l [n] }{d \lambda^{2} } \leq 0.
    \label{second_deriv}
\een
Note that all inequalities for $E\cw$, both coordinate-scaled (Eqs.~\ref{uc-ineq},~\ref{uc-ineq2}) and $\lambda$-dependent (Eqs.~\ref{hubbard_uc},~\ref{hubbard_uc2}), are also true for $U\cw=E\cw-T\cw$, the potential contribution to correlation. The HX energy (Eq.~\ref{high-dens-x}) may be extracted via

\ben
  E\hxw[\n] = \lim_{\lambda \rightarrow 0} E\l\hxcw [n]/\lambda.
    \label{hubbard_high-dens-x}
\een

 Our last condition concerns the relationship between DFT and traditional approaches to quantum chemistry.   In the ground state, it has long been known~\cite{crisostomo2022can,lowdin1955quantum} that $0 \geq E\c\HF \geq E\c$, where $E\c\HF$ is the traditional definition of the correlation energy, i.e., relative to the Hartree-Fock (HF) energy
(we treat only restricted HF here, RHF).  Given the complications of EDFT, we discuss here only the case of the first singlet bi-ensemble for two electrons.   In this case, we equate  EHF with an EDFT EXX calculation ('exact exchange only').  The only difference between EHF and EDFT is that the EHF quantities are evaluated on the approximate EHF density, while EDFT quantities are evaluated on the exact density.  Exactly the same variational reasoning leads us to
\ben
0 \geq E\cw\HF \geq E\cw[n]
\een
where $E\cw\HF = E\w -E\w\HF$, and $E\w\HF$ minimizes $F\w=T\sw +E\hxw$.  We leave the more general case to braver souls.

\begin{figure}[t]
    \centering
    \includegraphics[width=0.4\textwidth]{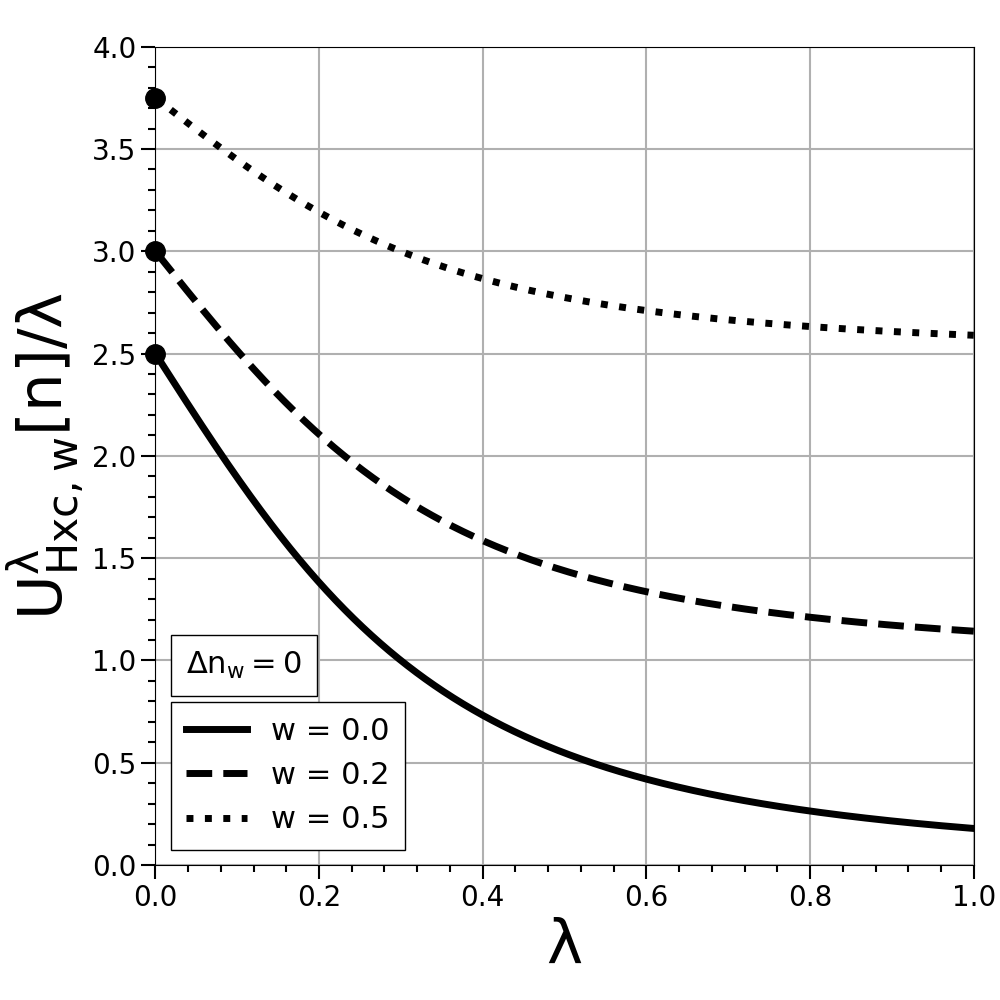}
    \caption{Ensemble adiabatic connection with $\Delta n\w = 0$ and $U = 5$; circles represent the weight-dependent HX energy, which the HXC expression approaches as $\lambda\to 0$ (Eq.~\ref{hubbard_high-dens-x}). More cases are provided in Fig. S8 of the supplemental material.}
    \label{fig:high-dens-x}
\end{figure}

The Hamiltonian of the Hubbard dimer is
\begin{equation}
    \hatH = -t\sum_{\sigma}(\hat{c}^{\dagger}_{1\sigma}\hat{c}_{2\sigma} + h.c.) + U\sum_{i}\hat{n}_{i\uparrow}\hat{n}_{i\downarrow} + \sum_{i}v_{i}\hat{n}_{i}, \\
    \label{hubbard_hamil}
\end{equation}
where $t$ is the hopping parameter, $U$ the on-site
electrostatic self-repulsion, and $v_{i}$ the on-site potential (which controls the asymmetry of the dimer). For this lattice system,  with $N=2$,
the electronic density is characterized by a single number, the difference
between occupations of the two sites, $\Delta n = n_2 - n_1$. 
The $\lambda$-dependence of any quantity is found by replacing $U$ by $\lambda U$, keeping $\Delta n$ fixed. We choose $t = 1/2$ everywhere.

We consider the simplest bi-ensemble, a mixture of the ground-state with the
first excited singlet. Full analytic expressions of $\ket{\Psi_0}$ and $\ket{\Psi_1}$,
as well as plots of various bi-ensemble quantities, are given in
the supplemental material in section 1. The value of $\Delta n\w$ is constrained
by $w$:

\ben
    |\Delta n\w| \leq 2\overline{w},
\een

where $\overline{w} = 1 - w$, i.e. is smaller than that of the ground state ($w=0$). Densities are shown in Fig. S1 of the supplemental material. The total energy of the ensemble is defined as
\begin{equation}
    E\w = \overline{w} \bra{\Psi_0}\hat{H}\ket{\Psi_0} + w \bra{\Psi_1}\hat{H}\ket{\Psi_1}.
    \label{hubbard_energy}
\end{equation}
Plots of $E\w$ are depicted in Figs. S2-S3 of the supplemental material, showing the quantity both as a function of $\Delta v$ and $\Delta n\w$. We also show analogous plots of $F\w = E\w - \Delta v \Delta n\w / 2$ in Figs. S4-S5. For this bi-ensemble, the exact HX energy has the simple analytical form \cite{deur2017exact}:
\begin{equation}
    E\hxw
    = \frac{U}{2}\left[1 + w + \frac{(1-3w)}{\overline{w}^2} \frac{\Delta n_w^2}{4}\right]. 
\end{equation}

\textbf{Inequalitites.}
We plot $\lambda$-dependencies in Fig.~\ref{fig:CorrScaling} that
have a definite sign according to Eq. \ref{uc-ineq}.  We show
several values of $w$ for two densities for $U=1$ (moderate correlation).  Scanning over all
$U$ and $\Delta n\w$, these inequalities are always satisfied. 
For the symmetric dimer ($\Delta n\w= 0$),  $w = 0.0$ has the largest maximum and 
$w = 0.5$ has the smallest. As $\Delta n\w$ is increased,
the trend disappears and the curves are not monotonic in $w$. 
For $w = 0.5$, the inequality becomes an equality 
for $\Delta n\w= 1$, the maximum representable value of $\Delta n$ for 
the ensemble.
In Fig.~\ref{fig:high-dens-x},
we show that all $U\hxcw$ curves approach their corresponding HX value
as $\lambda\rightarrow0$, in accordance with Eq.~\ref{hubbard_high-dens-x}. More plots of Figs.~\ref{fig:CorrScaling} and \ref{fig:high-dens-x} for various combinations of $w$ and $\Delta n\w$ are provided in Figs. S6-S8 of the supplemental material.

The non-monotonic behavior in Fig.~\ref{fig-1} can be easily understood. By definition, $E\w(\Delta v)$ is linear in $w$, as is $F\w$.  But, when converted to density 
functionals, and with KS quantities subtracted,
these become highly non-monotonic, as shown in
Figs. S2 and S3 in the supplemental material.

\begin{figure}[H]
    \centering
    \includegraphics[width=0.45\textwidth]{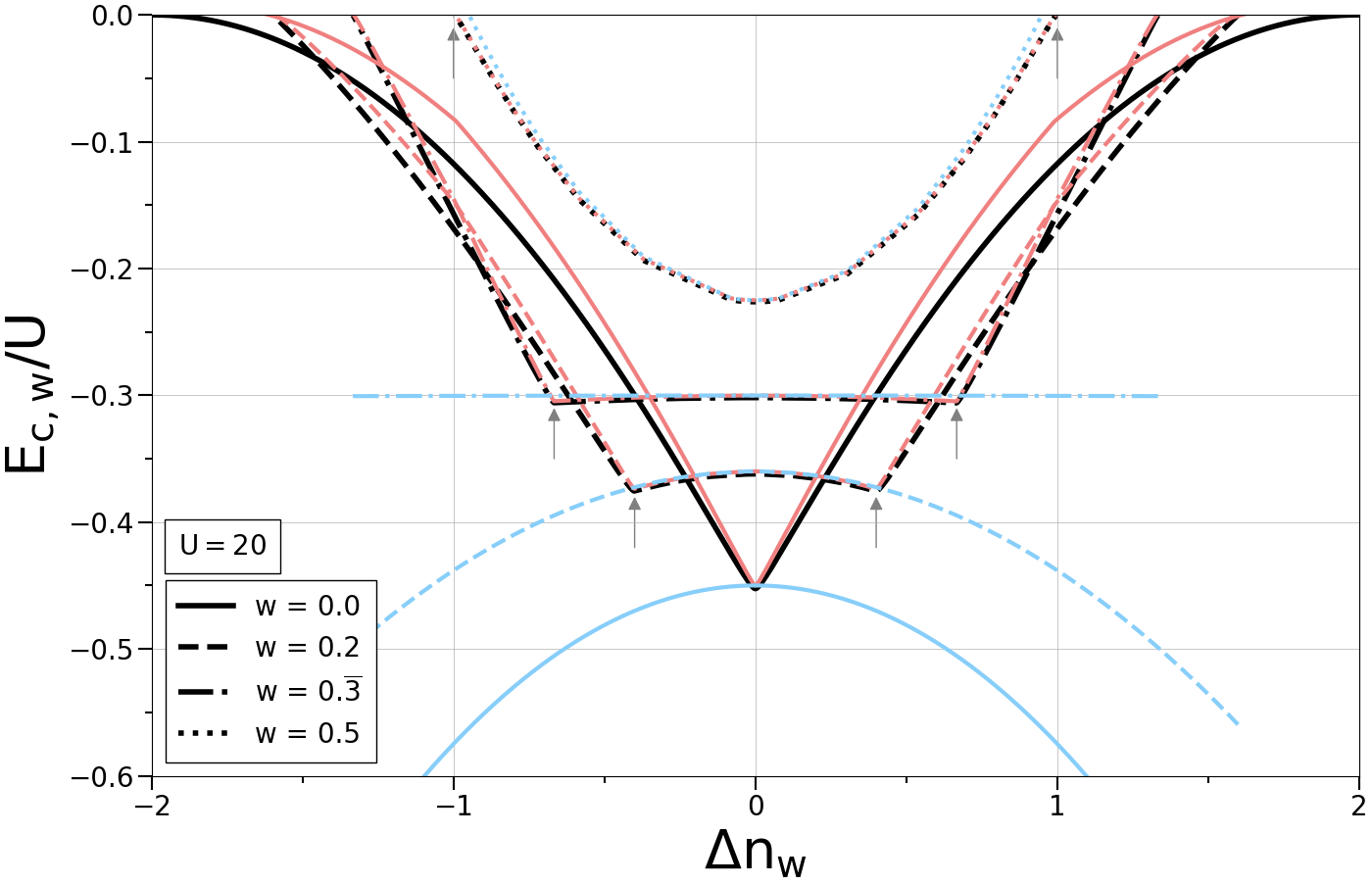}
    \caption{Exact correlation energy (black), leading-order expansion in large $U$ (red) and the expansion in the symmetric limit (blue) for the correlation energy are all plotted as a function of the exact density. Small arrows indicate the region between $2w$ and $-2w$ where the symmetric expansion matches the exact.}
    \label{fig:Ecw_Deur}
\end{figure}

\begin{figure*}
    \centering
    \includegraphics[width=\textwidth]{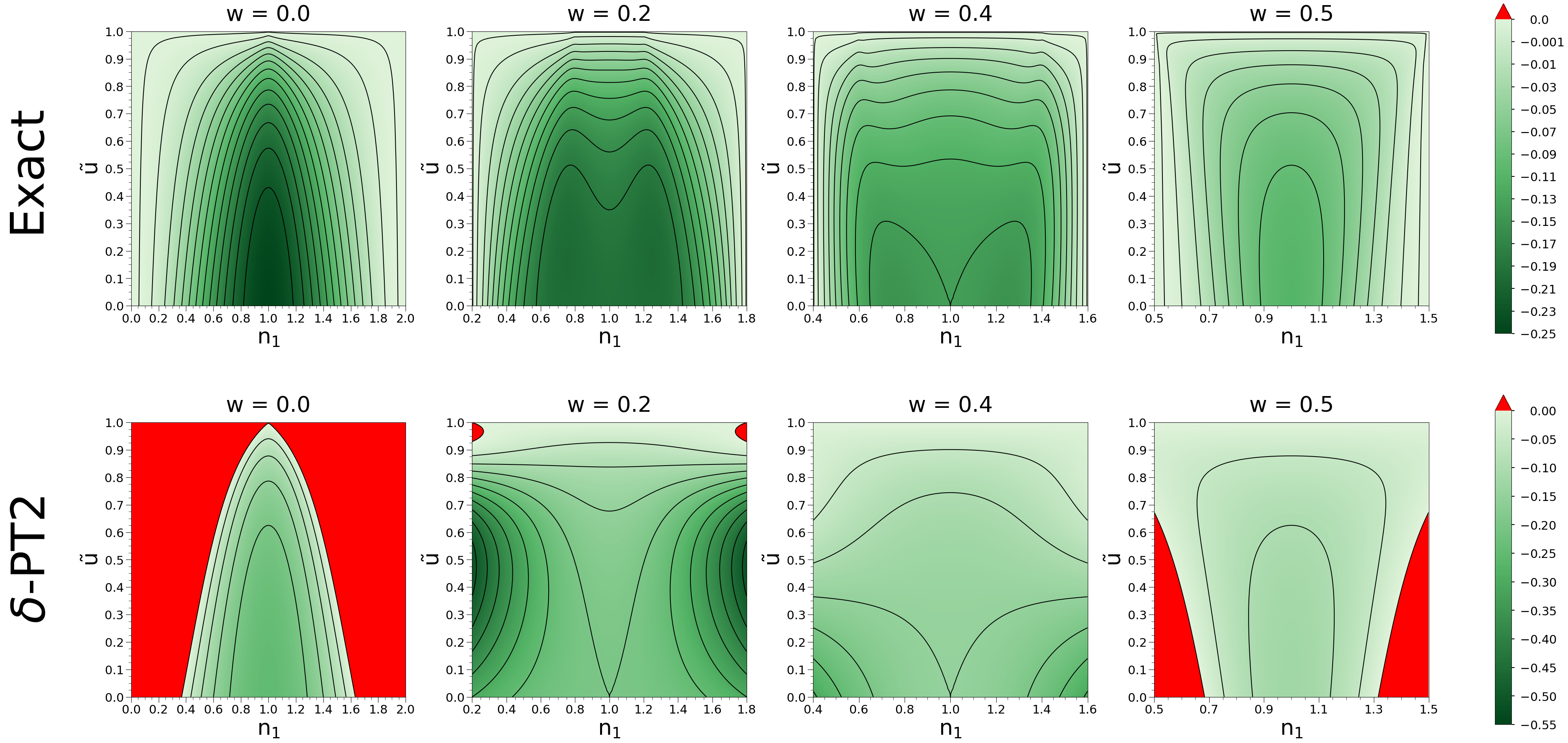}
    \caption{The second derivative of the Hubbard dimer bi-ensemble correlation energy with respect to $U$ for all values of the reduced variable $\tilde{u} = U/\sqrt{1 + U^2}$. The concavity condition is satisfied, but is violated by the $\delta$-PT2 approximation in certain regimes (red denotes positive). But $U$-PT2 automatically satisfies it, by construction.}
    \label{fig:contours}
\end{figure*}

\textbf{Strong Correlation.} In Fig.~\ref{fig:Ecw_Deur} we plot the exact correlation energy, our approximation (Eq. S.16 of the supplemental material), and the symmetric limit expansion of Deur {\em et al.} \cite{deur2018exploring}, each evaluated at the exact density.  The last yields the strongly correlated correlation energy only for $\vert \Delta n\w \vert \leq 2w$, but our expansion yields the correct limit for all allowed $\Delta n\w$, including the slope discontinuity at $|\Delta n\w| = 2w$. Such $w$-dependent derivative discontinuities occur only in EDFT. The approximate weight-dependent strongly correlated correlation energy is derived in the supplemental material along with further analysis of the energy components and approximation of the density. For the strong-interaction limit of the dimer, the correlation energy contains a non-trivial weight-dependence. This differs  from real space \cite{gould2023electronic} where the energies were found to be weight-independent. This is not a counter example, because the dimer is a site-model. This difference manifests in the expansion of the strongly correlated energies in powers of the coupling-constant.  Our first correction, relative to the leading term, is  $\mathcal{O}(\lambda^{-2})$ and differs qualitatively from the $\mathcal{O}(\lambda^{-1/2})$ behavior found by Gould and coauthors.

\textbf{Concavity Condition of the Correlation Energy.}
 We illustrate the concavity condition of Eq.~\ref{second_deriv} using contour plots depicting all possible combinations of $U$ and $n_1$, making use of the reduced variable $\tilde{u} = U/\sqrt{1 + U^2}$.
 Illustrated by Fig.~\ref{fig:contours}, the second derivative is negative for all values of $U$ and thus satisfies the concavity condition for all electronic correlation strengths.


 The standard use of exact conditions in DFT is to ensure that approximate functionals satisfy them.~\cite{pederson2023reassessing}   We illustrate our conditions by applying them to existing approximations on the Hubbard model.   The first is the standard many-body expansion in powers of the interaction, $U$, which we perform up to 2nd-order, i.e., the analog of M\o ller-Plesset perturbation theory, denoted $U$-PT2.   The second is less familiar:  an expansion in powers of $\Delta n$ around the symmetric case, $\Delta n=0$, called $\delta$-PT2.~\cite{deur2018exploring}
This can be considered a (tortured) analog of the gradient expansion of DFT, \cite{perdew1996generalized} as it is an expansion around the uniform limit.   Fig.~\ref{fig:contours} shows that the $\delta$-PT2 approximation violates the concavity condition, even for $w=0$, while $U$-PT2 never does.   The violations are not monotonic with increasing weights, as $w=0.4$ has none.  
Deur {\em et al.} reported that, compared to U-PT2, $\delta$-PT2 produced more accurate equi-ensemble energies and densities. Likely, the accuracy of $\delta$-PT2 could be further improved by imposing concavity.
Recent advances in EDFT, such as the direct ensemble correction \cite{yang2017direct} and the perturbative EDFT method, \cite{gould2022single} are explicitly computed in the perturbative limit, $w\rightarrow 0^{+}$.  If an approximation
is derived {\em before} such a limit is taken, and its ground-state approximation satisfies concavity; the resulting approximation should satisfy concavity also.

\textbf{Quantum Chemistry.} Finally,  we examine in detail the difference between the DFT and HF correlation energies and their components in Figs.~S5, S6, and S7 in the supplemental material. We provide plots of the exact/EHF total correlation energies for the dimer bi-ensemble, where we show the ground-state inequalities ($E\HF\c \geq E\c$) holds for any $w$-value (Figs.~S3 and S4). It also is known that $T\c\HF$ can become negative in the ground state of the Hubbard dimer,~\cite{giarrusso2022comparing} and we find this is also true when $w\neq 0$,  but this is likely an artifact of lattice Hamiltonians that cannot occur in the real-space analog.\cite{crisostomo2023seven,crisostomo2022can} 

This work provides new exact conditions for EDFT which can be used to analyze and/or improve new approximations in EDFT. Further work is being performed to improve approximations and provide a pathway to accurate EDFT functionals. 

In the supplemental material for this article we provide analytical expressions (and plots) for various weight-dependent quantities of interest for the Hubbard dimer bi-ensemble, giving both the exact/EHF solution.

T.R.S. thanks the Molecular Software Science Institute for funding, award no. 480718-19905B, and Sina Mostafanejad for fruitful discussions. K.B., J.K., and S.C. acknowledge support from NSF award number CHE-2154371. A.P.J. acknowledges financial support for this publication from Cottrell Scholar Award \#28281, sponsored by Research Corporation for Science Advancement.

\bibliography{main}

\label{page:end}

\end{document}


\sf
\coloredtitle{\TitleOfPaper}

\coloredauthor{Thais R. Scott}
\affiliation{Department of Chemistry, University of California, Irvine, CA 92697}

\coloredauthor{John Kozlowski}
\affiliation{Department of Chemistry, University of California, Irvine, CA 92697}

\coloredauthor{Steven Crisostomo}
\affiliation{Department of Physics \& Astronomy, University of California, Irvine, CA 92697}

\coloredauthor{Aurora  Pribram-Jones}
\affiliation{Department of Chemistry, University of California, Merced, CA 95343}

\coloredauthor{Kieron Burke}
\affiliation{Department of Chemistry, University of California, Irvine, CA 92697}
\affiliation{Department of Physics \& Astronomy, University of California, Irvine, CA 92697}

\date{\today}


\begin{abstract}
Below we list the analytical expressions defining the Hubbard dimer bi-ensemble, and plot various weight-dependent quantities of interest. In all plots we set $t = 1/2$.
\end{abstract}

\maketitle

\tableofcontents


\sec{Exact Solution}

We begin with the analytic solutions of the Hubbard dimer that were used to create a bi-ensemble of the ground and first-excited singlet states. Solving the dimer Hamiltonian (Eq.26), one obtains the energies:
\begin{equation}
    E_i = \frac{2U}{3} + \frac{2r}{3}\cos{\left[\theta + \frac{2\pi (i + 1)}{3}\right]}, \quad\quad i = 0,1.
\end{equation}
Here we have defined
\begin{equation}
\begin{gathered}
    r = \sqrt{3\left(4t^2 + \Delta v^2\right) + U^2}, \quad\quad \cos{\left(3\theta\right)} = \frac{9U\left(\Delta v^2 - 2t^2\right) - U^3}{r^3}, \notag
\end{gathered}
\end{equation}
where $t$ represents the hopping parameter, $U$ the on-site electrostatic self-repulsion, and $\Delta v = v_2 - v_1$ the on-site potential difference. Furthermore, the wavefunction of each state may be written as
\begin{equation}
    \ket{\Psi_i} = \alpha_i \left(\ket{12} + \ket{21}\right) + \beta_i^+ \ket{11} + \beta_i^- \ket{22},
    \label{HubbardWF}
\end{equation}
with coefficients:
\begin{equation}
\begin{gathered}
    \alpha_i = \frac{2t \left(E_i - U\right)}{c_i E_i}, \quad\quad \beta_i^\pm = \frac{U - E_i \pm \Delta v}{c_i}, \vspace{7pt} \\
    c_i = \sqrt{2 \left[\Delta v^2 + \Bigl(E_i - U\Bigr)^2 \Bigl(1 + 4t^2/E_i^2\Bigr)\right]}. \notag
\end{gathered}
\end{equation}
Here $\ket{ij}$ represents a state in which an electron is present at sites $i$ and $j$. These analytical expressions (both of the energy and wavefunction) may also be found in the appendices of References \citenum{deur2017exact} and \citenum{smith2016exact}.

\ssec{Densities}
We use the same bi-ensemble as described in the main text; the two lowest lying singlet states in the Hubbard dimer. The density of each state is trivially $\Delta n_i = 2 \big[\big(\beta_i^-\big)^2 - \big(\beta_i^+\big)^2\big]$, meaning that the ensemble density is
\begin{equation}
    \Delta n\w = 2 \wb\bigg[\big(\beta_0^-\big)^2 - \big(\beta_0^+\big)^2\bigg] + 2 w\bigg[\big(\beta_1^-\big)^2 - \big(\beta_1^+\big)^2\bigg],
    \label{coeffs}
\end{equation}
where $\wb = 1 - w$. Plots of this difference are included below in Fig.~\ref{fig:DnVsDv} for three interaction strengths, $U=0,\,1,\,\&\,5$.

\begin{figure}[ht!]
    \centering
    \includegraphics[width=0.48\textwidth]{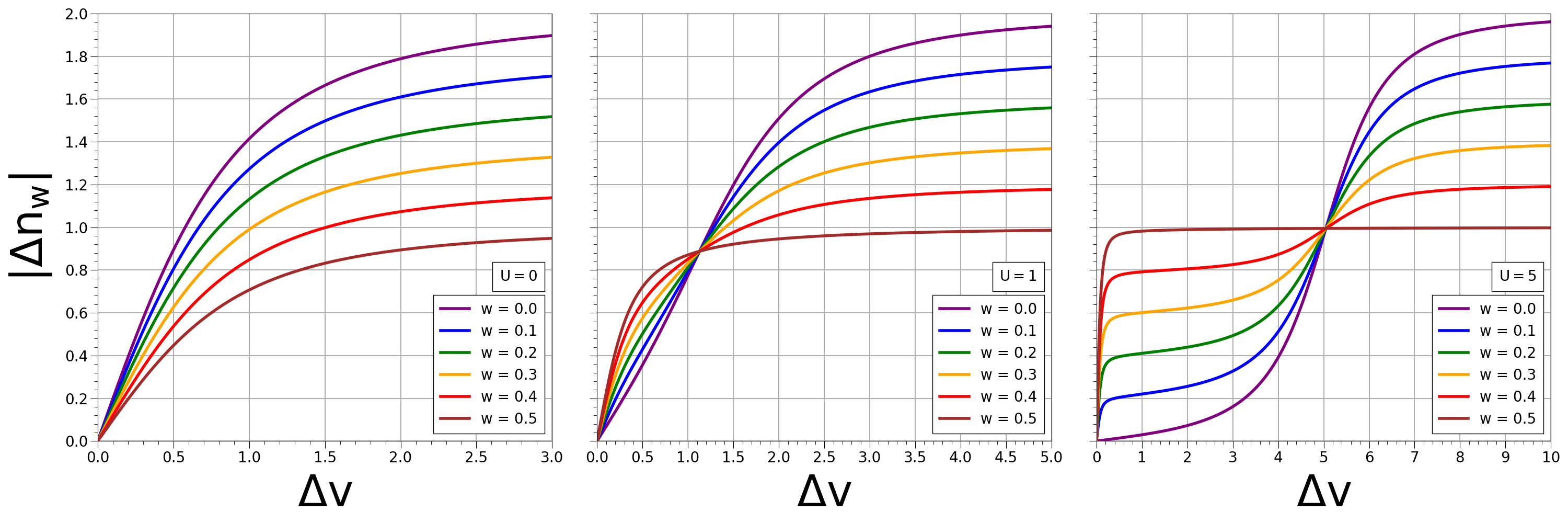}
    \caption{Absolute value of the density of the Hubbard dimer bi-ensemble, plotted for various weight values as a function of $\Delta v$. Here we set $U=0$ (left), $U=1$ (center), and $U=5$ (right).}
    \label{fig:DnVsDv}
\end{figure}    

Analyzing this figure, it is clear that there are vast differences in the behavior of $\Delta n\w$ with respect to the value of $U$, illustrating the importance of developing weight-dependent approximations for electronic correlation. Two characteristics are present in all plots of $\Delta n\w$, these being an adherence to the symmetric limit ($\Delta v = \Delta n\w = 0$) and a maximum value constraint imposed by the representability condition $|\Delta n\w| \leq 2\wb$, which each curve approaches as $\Delta v \to \infty$.

As predicted by EDFT (eq X), the density is linear in $w$. For sufficiently large $U$, there exists a value of $\Delta v$ at which the initial slope of the first excited state density difference is negative, but it always becomes positive for sufficiently large $\Delta v$.  Thus there is a specific value of $\Delta v$ at which the first excited state density vanishes, and all curves meet at that point, independent of $w$.  This point tends to $\Delta v=U$ as $U$ becomes large.

\begin{figure}[htb!]
    \centering
    \includegraphics[width=0.35\textwidth]{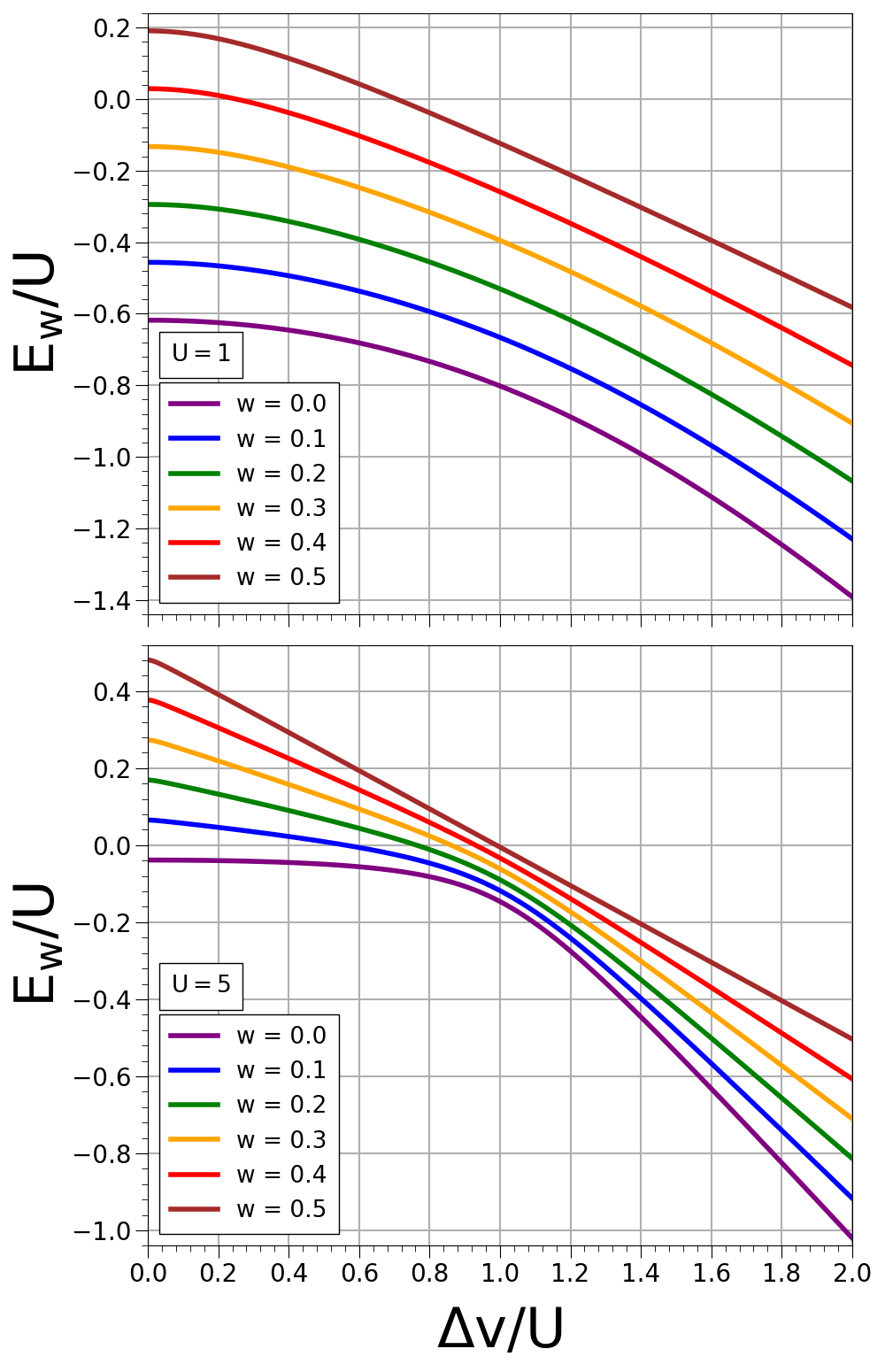}
    \caption{Total energy of the Hubbard dimer bi-ensemble plotted as a function of $\Delta v$ for various $w$ values. Here we set $U=1$ (top) and $U=5$ (bottom).}
    \label{fig:EnVsDv}
\end{figure}

\begin{figure}[htb!]
    \centering
    \includegraphics[width=0.35\textwidth]{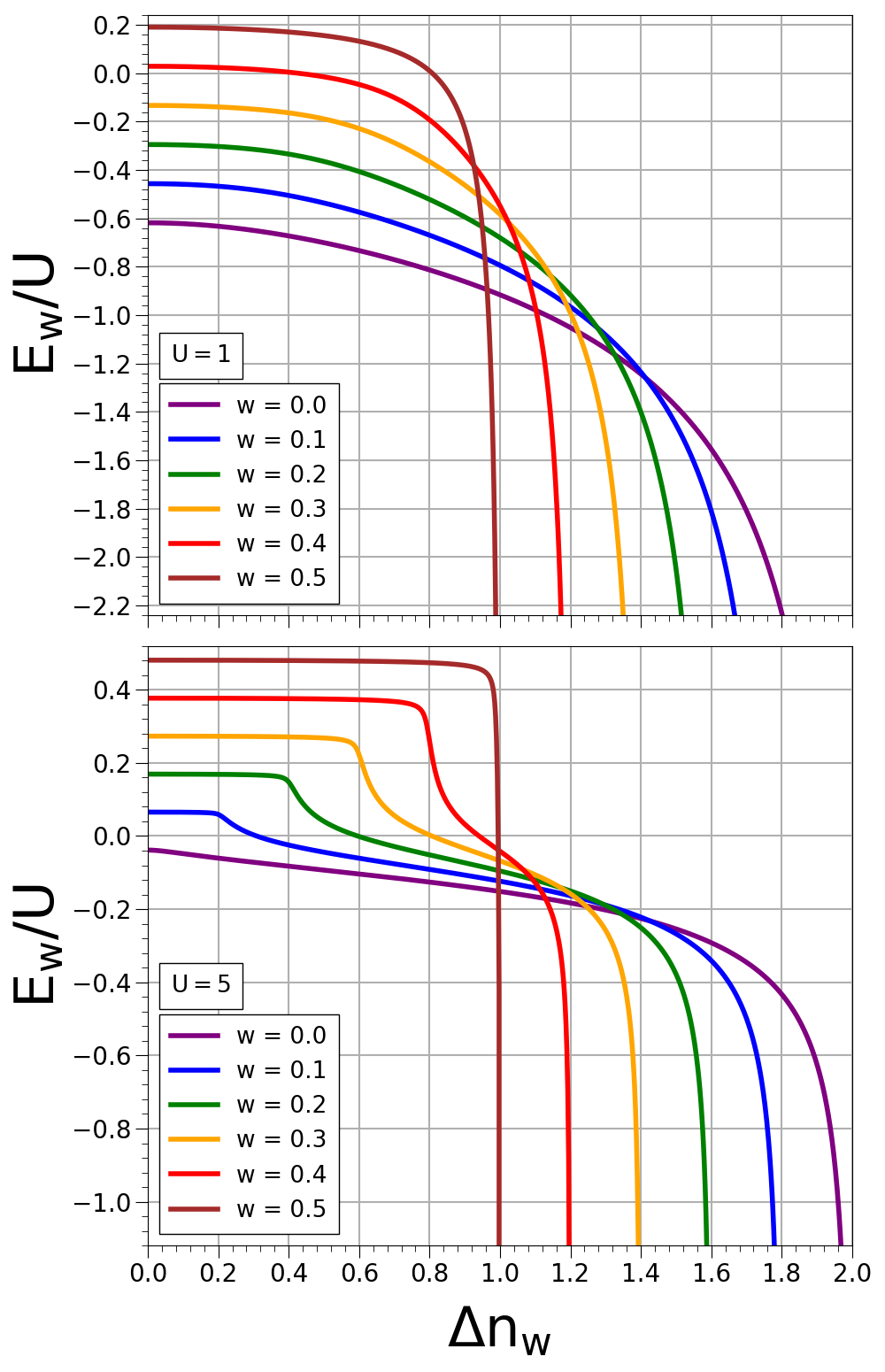}
    \caption{Total energy of the Hubbard dimer bi-ensemble plotted as a function of $\Delta n\w$ for various $w$ values. Here we set $U=1$ (top) and $U=5$ (bottom).}
    \label{fig:EnVsDn}
\end{figure}

For finite values of $U$, a trend in curve steepness with respect to the weight is evident for small $\Delta v$; the steepness of each curve is directly proportional to the value of $w$, signifying that ensembles with larger $w$ values approach their maximum value more quickly. The severity of steepness increases drastically as $U$ is increased, as shown by the behavior of the $U=5$ curves as $\Delta v \to 0$. Here, the densities increase rapidly to $|\Delta n\w| \approx 2w$, becoming nearly perfectly anti-symmetric around $\Delta v = U$, with a very sharp dive to $0$ for very small $\Delta v$.It also appears that all $\Delta n\w$ curves approach the same value at $\Delta v \approx U$ as $U \to \infty$. As $U\rightarrow\infty$ the density forms a step function, flipping from $2w$ to $2\wb$ at $\Delta v = U$ (see Fig. \ref{fig:LO_site_occs}).

\begin{figure}[ht!]
    \centering
    \includegraphics[width=0.35\textwidth]{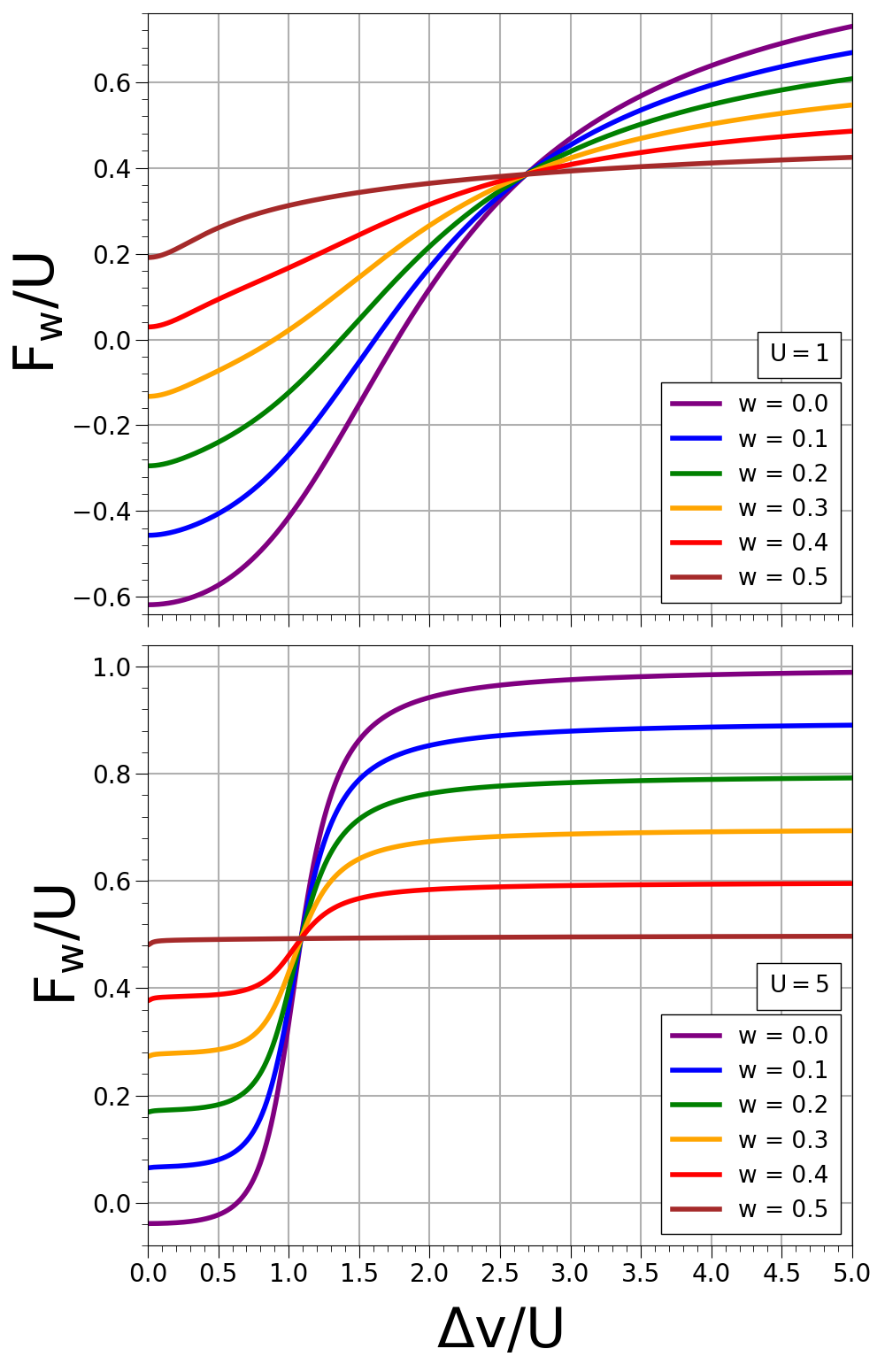}
    \caption{$F\w / U$ of the Hubbard dimer bi-ensemble plotted as a function of $\Delta v$ for various $w$ values. Here we set $U=1$ (top) and $U=5$ (bottom).}
    \label{fig:FwVsDv}
\end{figure}

\begin{figure}[htb!]
    \centering
    \includegraphics[width=0.35\textwidth]{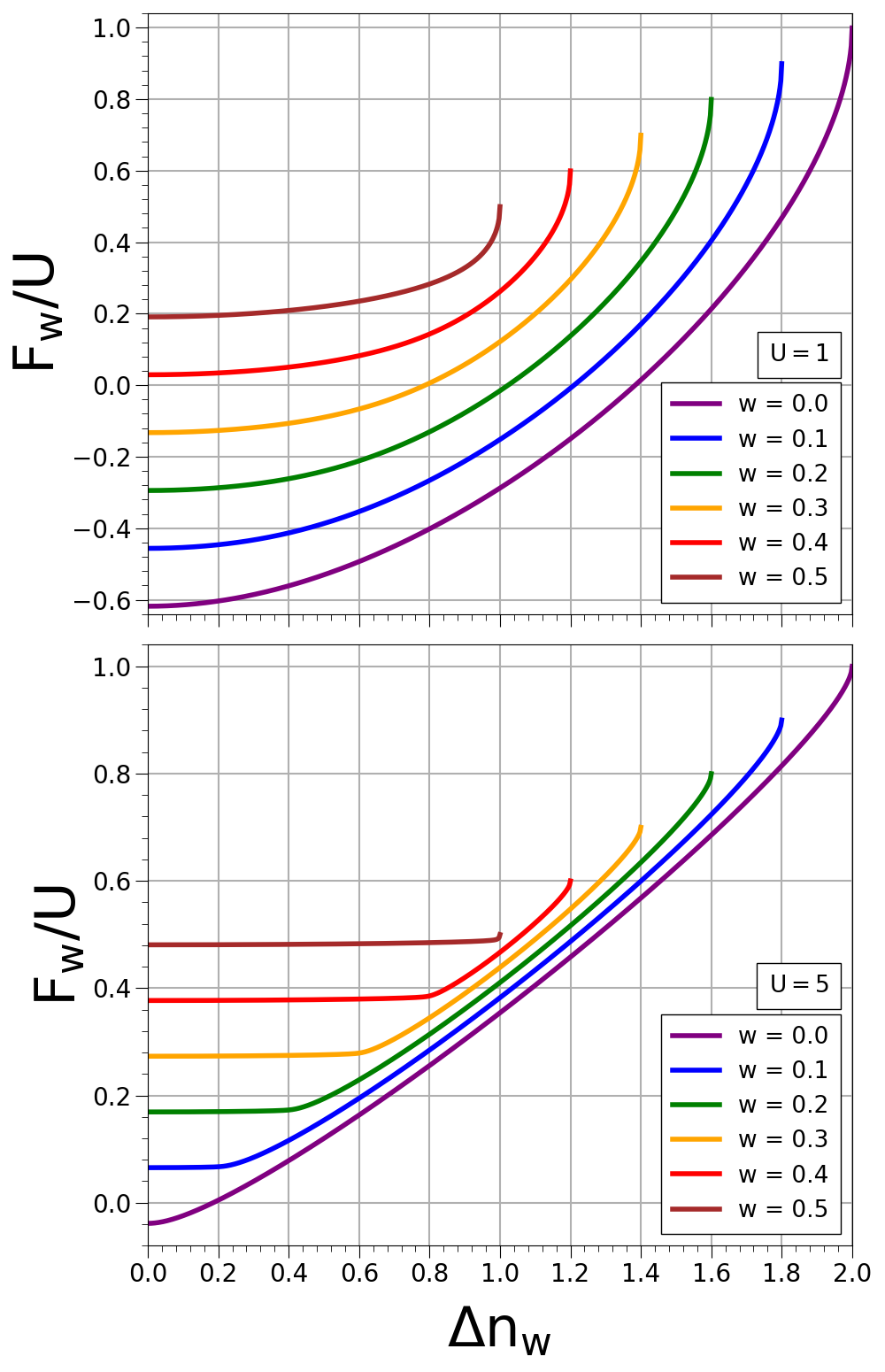}
    \caption{$F\w / U$ of the Hubbard dimer bi-ensemble plotted as a function of $\Delta n\w$ for various $w$ values. Here we set $U=1$ (top) and $U=5$ (bottom).}
    \label{fig:FwVsDn}
\end{figure}

\ssec{Total Energies}

In this section we depict plots of the total bi-ensemble energy, $E\w$, defined as the weighted sum of Eq.28. Here we choose to depict two interaction strengths ($U=1$ and $U=5$), plotting both as a function of $\Delta v$ and $\Delta n\w$ below. We make use of the derivation put forth by Deur {\em et al.}~\cite{deur2017exact} to define:
\begin{equation}
    T_{s,w} = -2t\sqrt{\wb^{2} - \Delta n\w^2 / 4} \label{eq:Tsw}
\end{equation}
\begin{equation}
    E_{Hx,w} = \frac{U}{2}\left(1 + w - \frac{(3w-1)}{\wb^{2}}\frac{\Delta n\w^{2}}{4}\right). \label{eq:EHXw}
\end{equation}
The correlation energies are then found by using the exact expressions:
\begin{equation}
    T_{c,w} = T_w - T_{s,w},
\end{equation}
\begin{equation}
    U_{c,w} = V\eew - E_{Hx,w},
\end{equation}
\begin{equation}
    E_{c,w} = T_{c,w} + U_{c,w}. \label{eq:corrE}
\end{equation}

For fixed $\Delta v$, 
Fig.~\ref{fig:EnVsDv} illustrates that $E\w$ is correctly linear in $w$.  The curves are rather boring for $U=1$, but develop a pinch around $\Delta v=U$ as $U$ grows larger (see Sec X).   However, 
Fig.~\ref{fig:EnVsDn} shows that, as a functional
of $\Delta n$, the curves are no longer linear in 
$w$.   They are not even monotonic.  Moreover
as $\Delta n\w \to 2\wb$, $E\w \to -\infty$, ensuring the curves cross. 
There exists interesting behavior as $U$ is increased, with $E\w$ becoming ever more slowly varying with
density for $\Delta n\w < 2w$. This behavior may be explained through the relationship between $\Delta n\w$ and $\Delta v$ shown in the right panel of Fig.~\ref{fig:DnVsDv}, where there is drastic change in $\Delta n\w$ for $\Delta v \approx 0$.

We also examine the properties of the Hubbard dimer equivalent of the universal part of the density functional, $F\w = E\w - \Delta v \Delta n\w / 2$, plotting again as a function of $\Delta v$ and $\Delta n\w$ in Figs.~\ref{fig:FwVsDv} and~\ref{fig:FwVsDn} respectively. In Fig.~\ref{fig:FwVsDv}, one can see that $F\w$ is linear with respect to bi-ensemble weight when plotted as a function of $\Delta v$. This characteristic is not present in Fig.~\ref{fig:FwVsDn}, where monotonicity is broken as $\Delta n\w \to 2\wb$. Furthermore, as $U$ is increased, one can see the appearance of regimes around $\Delta n\w = 2w$, with $F\w$ being nearly independent of $\Delta n\w$ for $\Delta n\w < 2w$ and linearly increasing as $\Delta n\w > 2w$. Additionally, the curves depicting $F\w$ tend to flatten as $w$ increases.

\begin{figure}[ht!]
    \centering
    \phantom{p}\newline\phantom{p}\newline
    \includegraphics[width=0.35\textwidth]{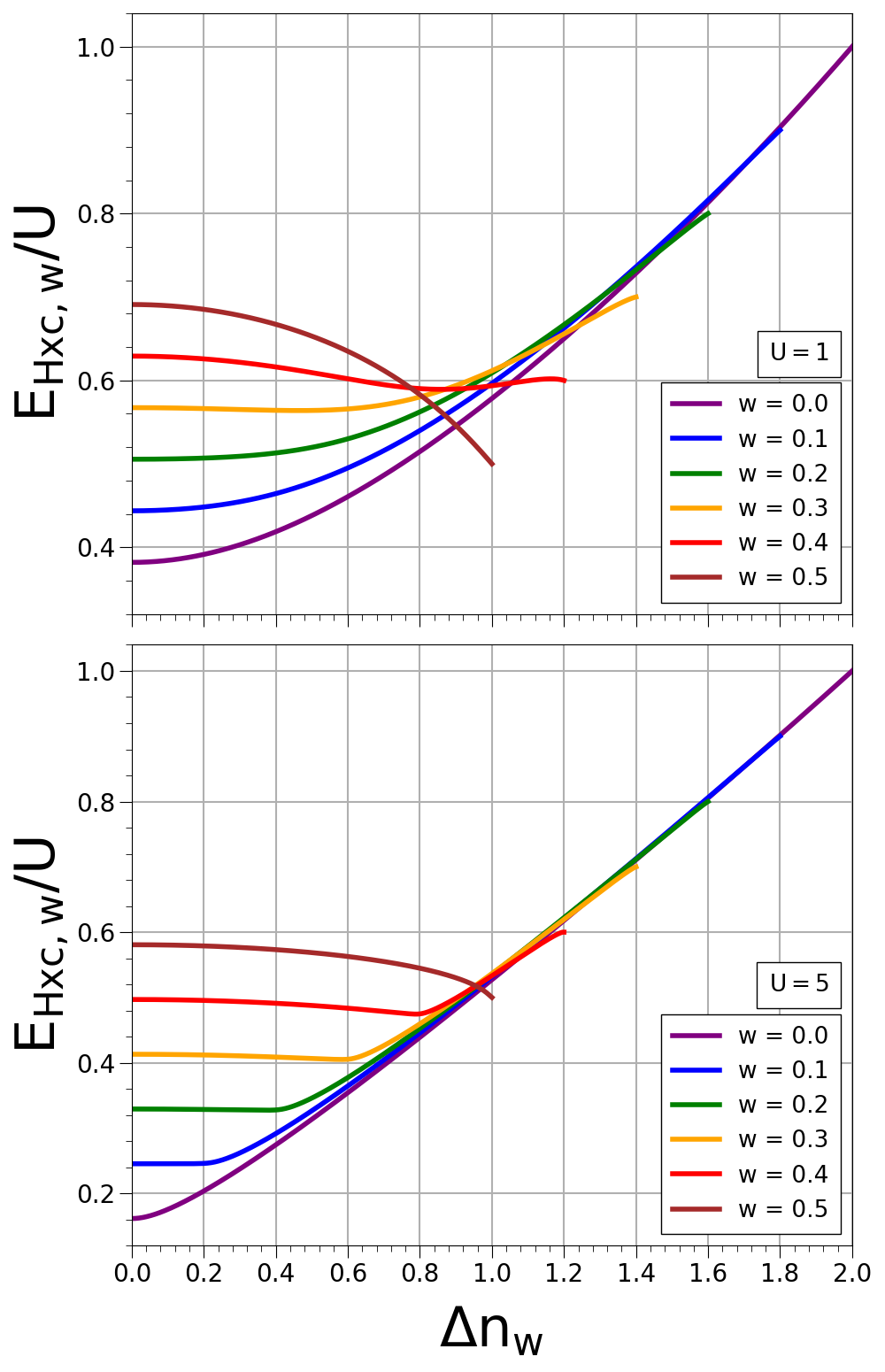}
    \caption{$E\hxcw /U$ of the Hubbard dimer bi-ensemble plotted as a function of $\Delta n\w$ for various $w$ values. Here we set $U=1$ (top) and $U=5$ (bottom).}
    \vspace{-22pt}
    \label{suppfig:ew}
\end{figure}

In contrast, we plot $E\hxcw$ in Fig.~\ref{suppfig:ew} and see it is non-monotonic in $w$. The curvature of $E\hxcw$ changes from convex to concave as the weight is increased and the $E\hxcw$ curves cross each other at various points; with the most noticeable crossings happening at $U=1$. However, the curves cross at all the values of $U$ plotted.

We conclude that curves become non-monotonic when plotted as a functional of the density instead of the potential, as the curves of Fig.~\ref{fig:DnVsDv} are definitely not monotonic.

\newpage
\ssec{Correlation Inequalities}

\begin{figure}[ht!]
    \centering
    \includegraphics[width=0.48\textwidth]{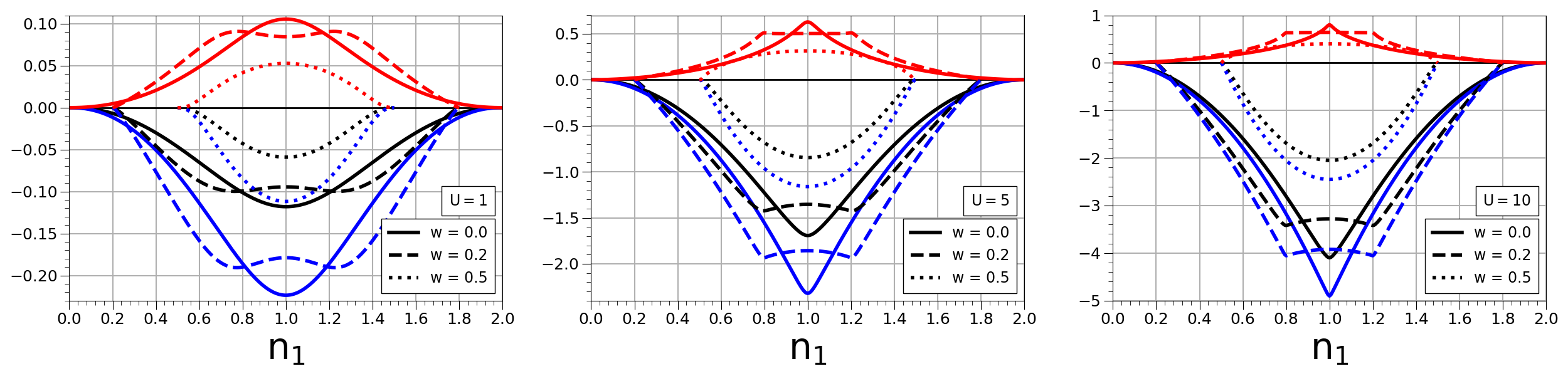}
    \caption{Variation of the potential (blue), kinetic (red), and total (black) correlation energies in the Hubbard dimer bi-ensemble, plotted as functions of site-occupation of the first site for various weights. We set $U=1$ in the left, $U=5$ in the center, and $U=10$ in the right.}
    \label{fig:CorrVsn}
\end{figure}

In this section we depict the correlation inequalities (Eq.21), showing more cases of Fig.2 of the main text. We highlight in Fig.~\ref{fig:CorrVsn} the definite signs of the correlation energy and its components. This validates results initially introduced by Pribram-Jones {\em et al.} \cite{pribram2014excitations}

\begin{figure}[ht!]
    \centering
    \includegraphics[width=0.48\textwidth]{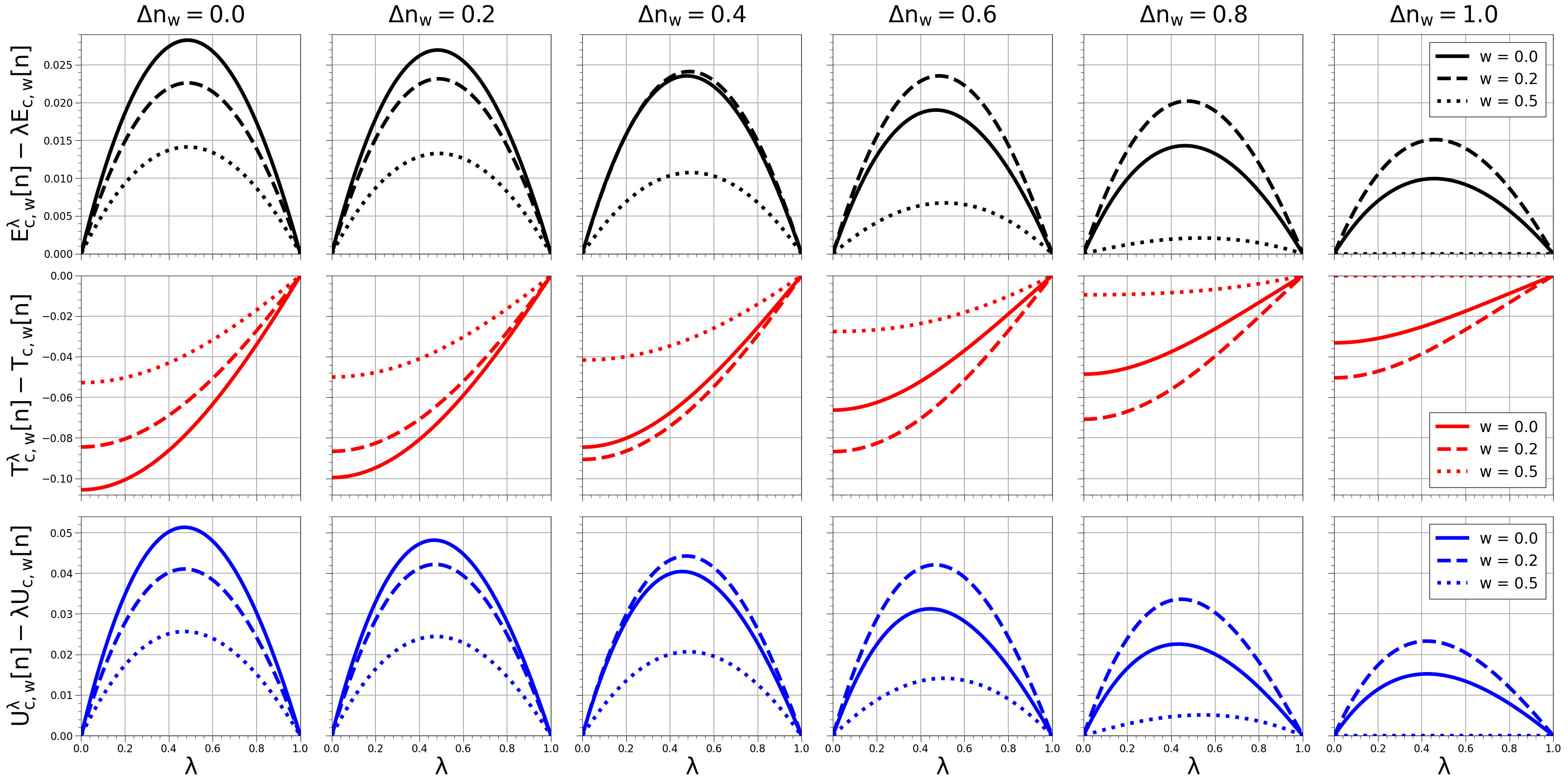}
    \caption{Correlation inequalities (Eq.21) for the total (top), kinetic (middle), and potential (bottom) correlation energies, depicted by varying $\lambda$ in the Hubbard dimer bi-ensemble with $U = 1$.}
    \label{fig:CsAllU1}
\end{figure}

\begin{figure}[ht!]
    \centering
    \includegraphics[width=0.48\textwidth]{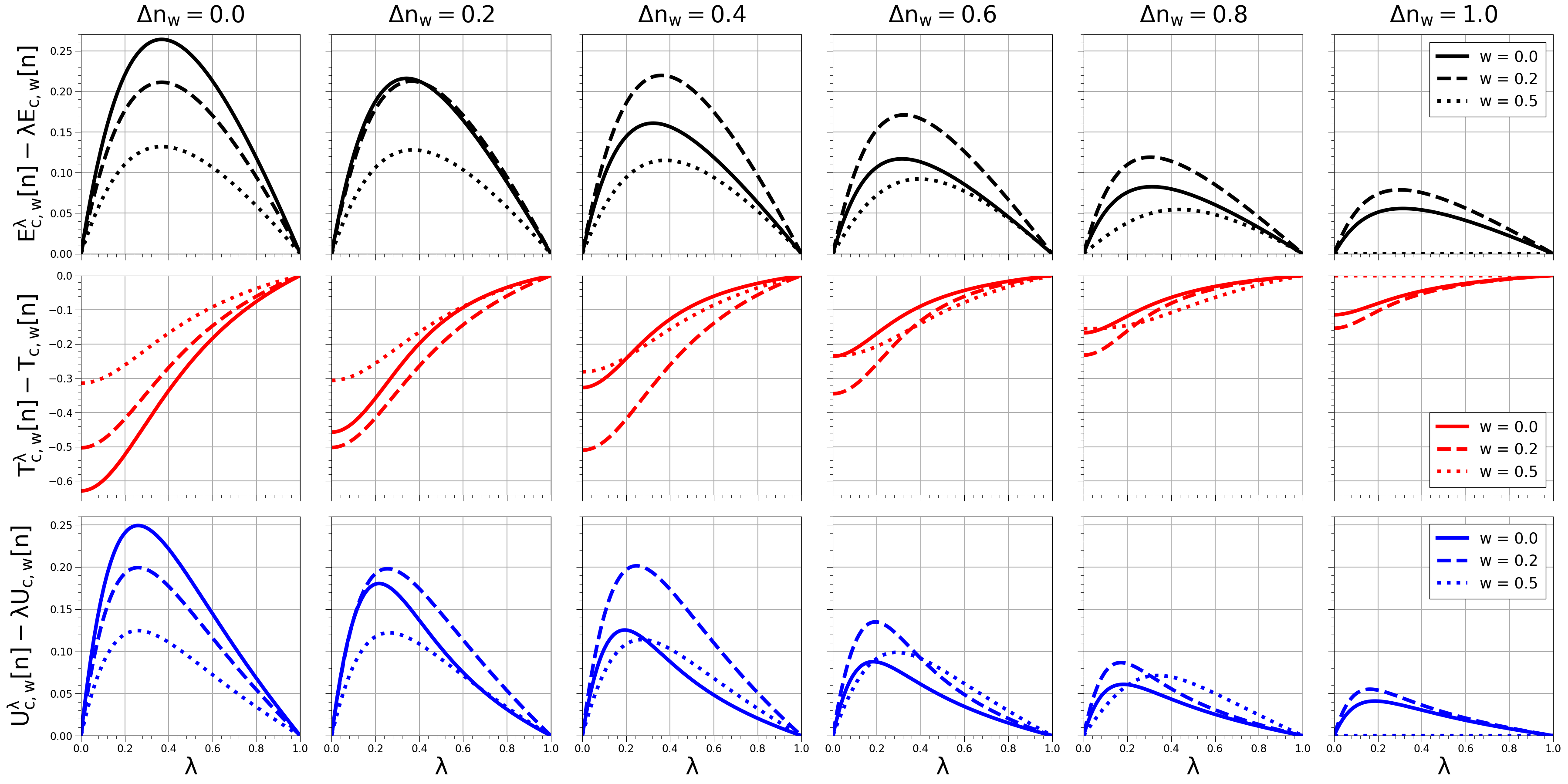}
    \caption{Correlation inequalities (Eq.21) for the total (top), kinetic (middle), and potential (bottom) correlation energies, depicted by varying $\lambda$ in the Hubbard dimer bi-ensemble with $U = 5$.}
    \label{fig:CsAllU5}
\end{figure}

Looking at Figs.~\ref{fig:CsAllU1} and~\ref{fig:CsAllU5}, one can see that the correlation inequalities of Eq.13 are satisfied for all values of $\Delta n\w$. As noted in the main text, there exists a clear trend with respect to weight for the symmetric dimer, with the ground-state having the maximum magnitude in each plot, and decreasing in magnitude with $w$. This trend no longer holds for any nonzero value of $\Delta n\w$. Alternative approaches were implemented in which $\Delta v$ was held fixed, again showing no clear trend for asymmetric dimers. Furthermore, it is clear that the inequalities of Eq.13 become equalities as $\Delta n\w \to 2\wb$, explaining the flat behavior of the $w=0.5$ curves for $\Delta n\w = 1$. It also appears that increasing $U$ morphs the shape of each curve, breaking symmetry around $\lambda = 0.5$ for $E\c$ and $U\c$.

\newpage
\ssec{Adiabatic Connection}

\begin{figure}[ht!]
    \centering
    \includegraphics[width=0.48\textwidth]{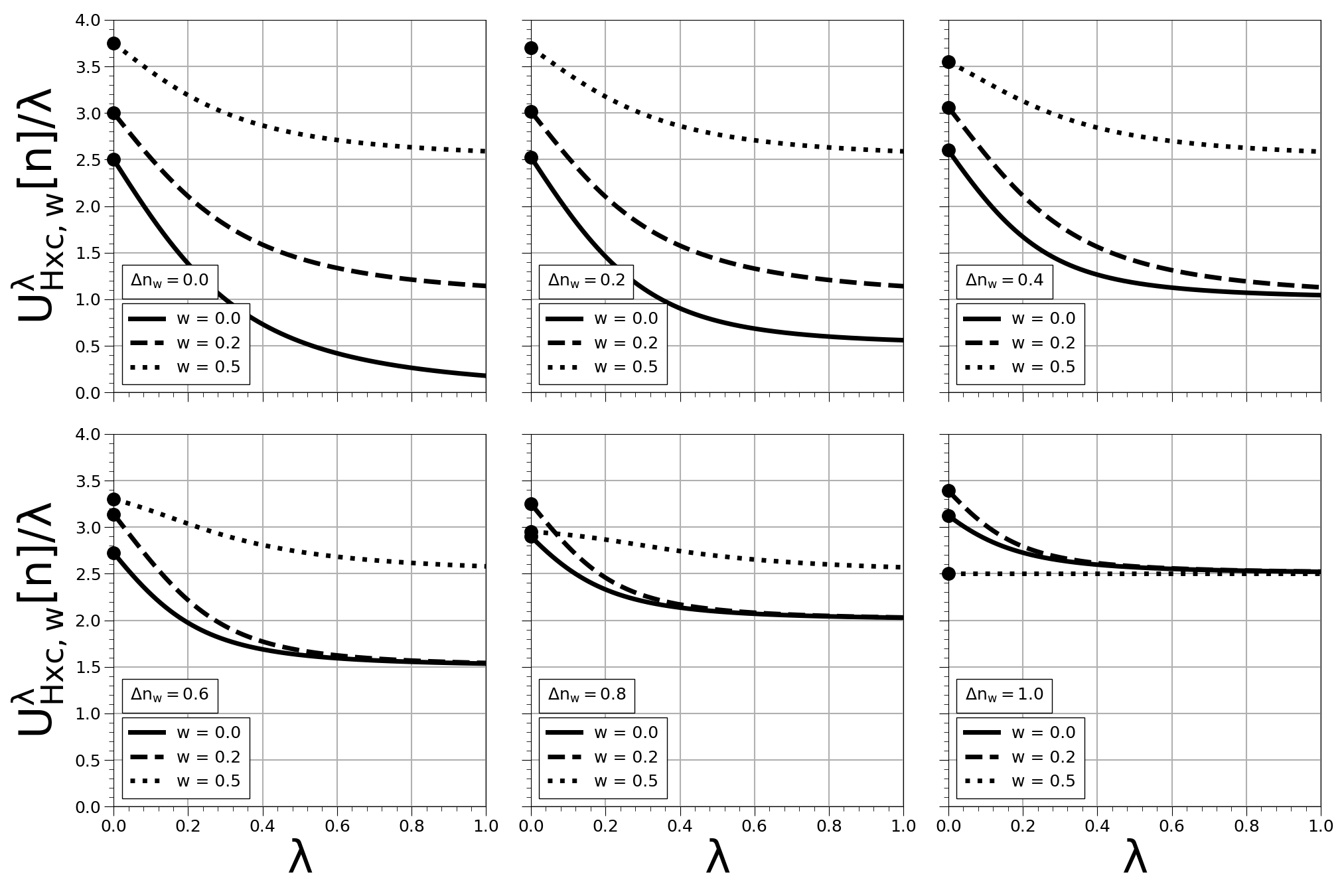}
    \caption{Ensemble adiabatic connection for $U = 5$ and various $\Delta n\w$; circles represent the weight-dependent HX energy, which the HXC expression approaches as $\lambda\to 0$.}
    \label{fig:CsExAll}
\end{figure}

In this section we include plots depicting more adiabatic connection curves than just that of Fig.3 of the main text.

We first note that all curves of a given weight look similar to ground-state DFT curves.  They are monotonically decreasing and are convex.
Looking at Fig.~\ref{fig:CsExAll} one can observe an interesting change in ordering of the HX energy values based on the weights. For $\Delta n\w < 0.6 $ the HX energy monotonicly decreases in value as the weights increase and the spacing between the values shrinks. However, after that point the ordering shifts and the $w=0.5$ weight has the maximum HX value. Additionally, as the $\Delta n\w \to 1$ the asymptotic value of the HXC expression is same for all weights.

\clearpage
\sec{Strong Correlation Limits of Energy Components}

\ssec{Large \textit{U} Expansions}

For fixed $w$, as $U$ becomes large, one can either keep $\Delta v$ fixed, or $\Delta v/U$ fixed. The former was explored by Deur et al. \cite{deur2018exploring}, and produces the pale blue curves of Fig.~5 of the main text. As is clear from the figure, the blue curves yield the correct answer only for $\vert \Delta n\w \vert \leq 2w$, which shrinks to a point as $w \to 0$.

The appropriate expansion to find $E\cw(\Delta n)$ for large $U$ is a different one. We take $U \to \infty$, but keep $\Delta v/U$ fixed. This must be done to include values of $\Delta n$ away from $\Delta n \approx 0$, while including all allowed values of $\Delta n$. A careful expansion yields the total energy as a function of $x = \Delta v /U $:
\ben E\w(x) \to U\left( g\w^{(0)}(x) + \frac{g\w^{(2)}(x)}{U^{2}} + \dots \right), \een
where

\ben g\w^{(0)}(x) = \frac{1}{3} \left(2-(c -\sqrt{3}  s)h\right)-\frac{2  s h w}{\sqrt{3}}, \label{eq:EcwSCLO}\een
and 

\ben
     g\w^{(2)}(x) =  \frac{1}{2h}\left(\frac{(\alpha -3) s}{\sqrt{3}} - (\alpha +1) c\right) + \frac{1}{h}\left(\alpha  c+\sqrt{3} s\right)w, \label{eq:gw2}\een
where $\alpha = \vert 4 x /(x^{2}-1)\vert$, $c = \cos(\phi)$, and $s=\sin(\phi)$ with,
\ben \phi = \frac{1}{3} \cos ^{-1}\left(\frac{3h^2-4}{h^{3}}\right), \quad \quad \quad h = \sqrt{3x^{2}+1}. \een
The angle $\phi$ is positive for all values of $x$, where it takes its maximal value of $\pi/3 $ as $x\to 0$  and minimal value of $0$ as $x\to \pm 1$, and in the limit $\phi(x \to \pm \infty) = \pi/6$. Because the angle is constrained to $0 \leq \phi \leq \pi/3$, the sine and cosine must always satisfy $0 \leq c,s \leq \sqrt{3}/2$.
 
From this we have the corresponding density via the exact expression $\Delta n\w = 2 dE\w/d(\Delta v)$,
\ben 
\Delta n\w(x) = 2\left(g\w^{(0)\prime}(x) + \frac{g^{(2)\prime}\w(x)}{U^{2}} + \dots\right), 
\een
where primes denote derivatives with respect to $x$.
Retaining only zero-order terms yields,

\ben
g\w^{(0)\prime}(x) = \frac{1}{h}\left(\frac{(\gamma -3 x)s}{\sqrt{3}}-(\gamma +x)c\right) + \frac{2}{h} \left( \gamma c +\sqrt{3}x s \right)w, \label{eq:LOn}
\een
with $\gamma = \sign(x(1-x^{2}))$.
Because the expansion in $U$ is singular near $x=0$ and $x = \pm 1$, $g\w^{(2)}(x)$ diverges at $\vert x \vert = 1$, and even $n\w^{(0)}(x) = 2g\w^{(0)\prime}(x)$ contains discontinuous steps. While formally correct in the limit $U \to \infty$, the exact density cannot have such steps, due to the Hohenberg-Kohn theorems. We therefore smooth Eq.~\ref{eq:LOn} with exponentials that become infinitely sharp as $U\to \infty$:

\ben \Delta n\w \approx \frac{x}{|x|} (f( | x| )-1) \left(1+(1-2w) \tanh \left(\frac{ \beta (| x| -1)}{2}\right)\right), \label{eq:SmoothLOn} \een
where $f(x) = \left(\exp(\beta x) +1\right)^{-1}$ is the Fermi-Dirac distribution with $\beta = 5U$. This is plotted in Fig.~\ref{fig:LO_site_occs} and is compared with the exact density. As $U \to \infty$, Eq.~\ref{eq:SmoothLOn} matches Eq.~\ref{eq:LOn}.

\begin{figure}
    \centering
    \includegraphics[width=0.48\textwidth]{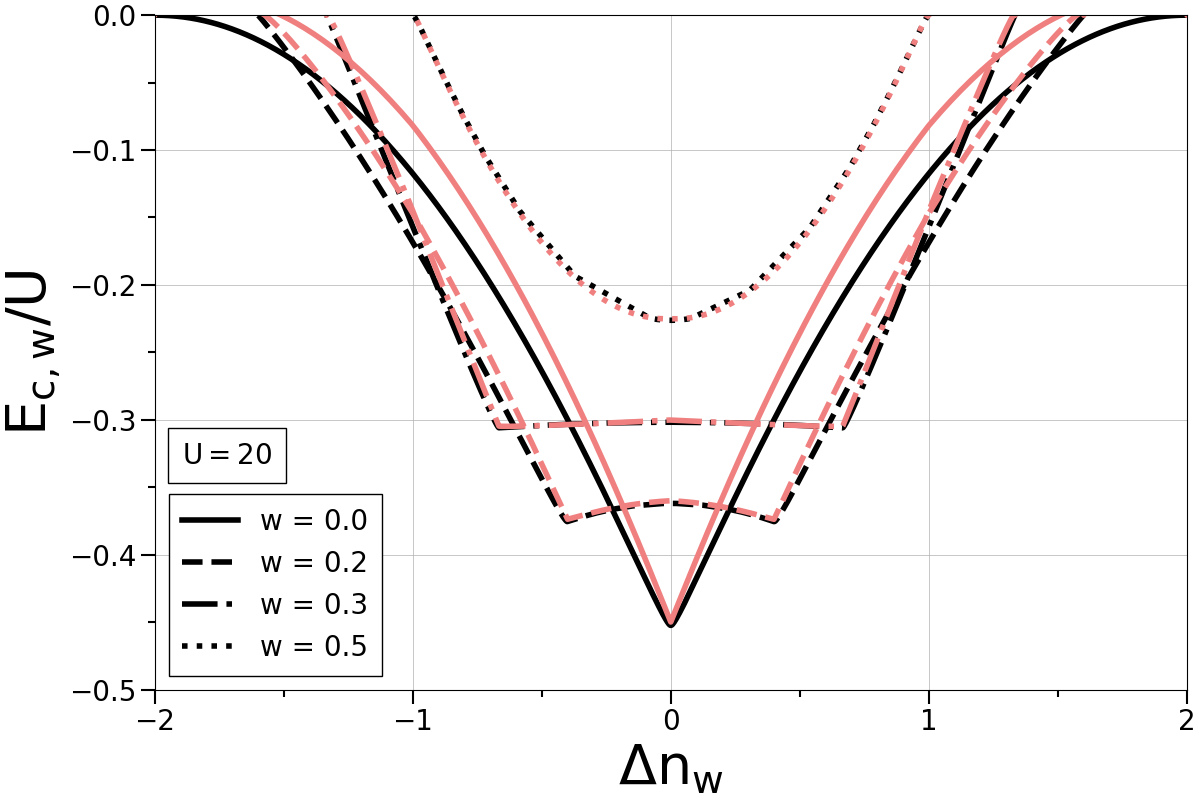}
    \caption{The exact (black) correlation energy as a function of the exact density and the leading-order expansion in large $U$ (light red) for the correlation energy, using \ref{eq:EcwSCLO}, plotted as a function of the smooth approximation for the density, \ref{eq:SmoothLOn}.}
    \label{fig:Ecw_smooth}
\end{figure}

\begin{figure}
    \centering
    \includegraphics[width=0.48\textwidth]{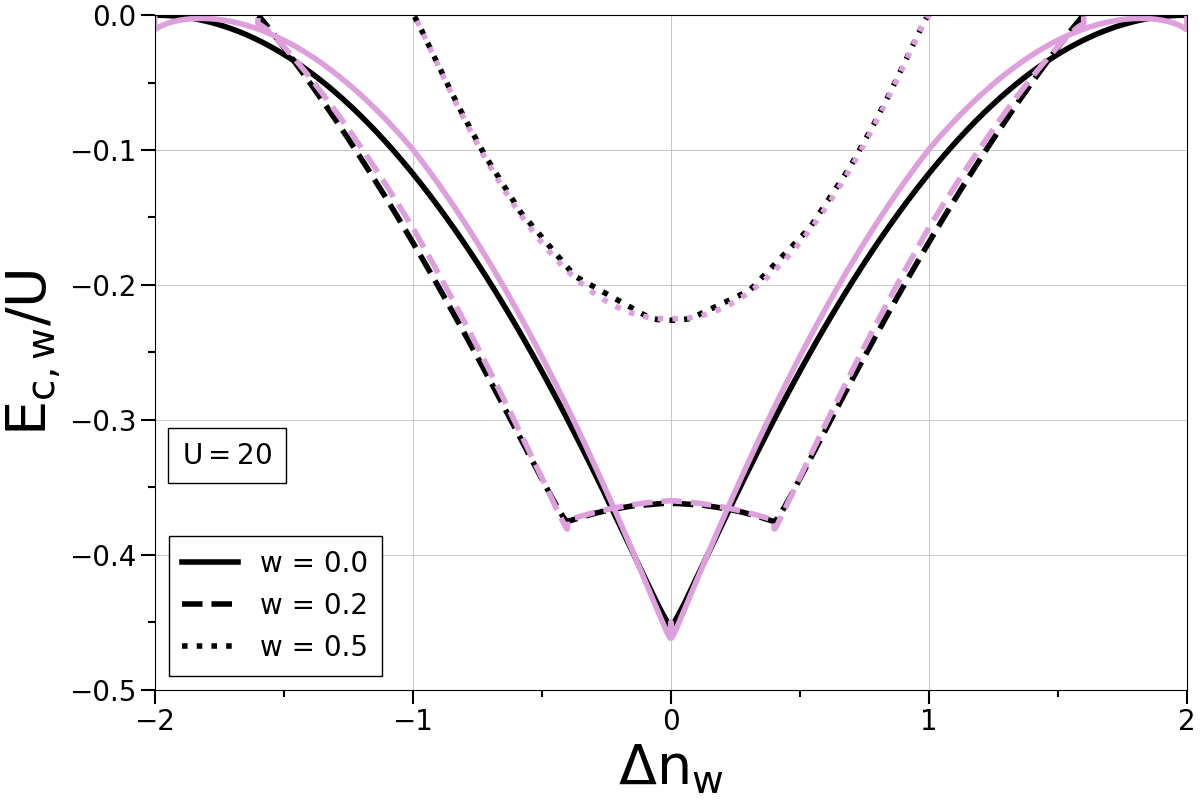}
    \caption{The exact (black) correlation energy as a function of the exact density and our higher-order expansion in large $U$ (light purple) for the correlation energy, \ref{eq:EcwSC}, plotted as a function of the smooth approximation for the density, \ref{eq:SmoothLOn}.}
    \label{fig:Ecw_smooth_HO}
\end{figure}

\begin{figure}
    \centering
    \includegraphics[width=0.48\textwidth]{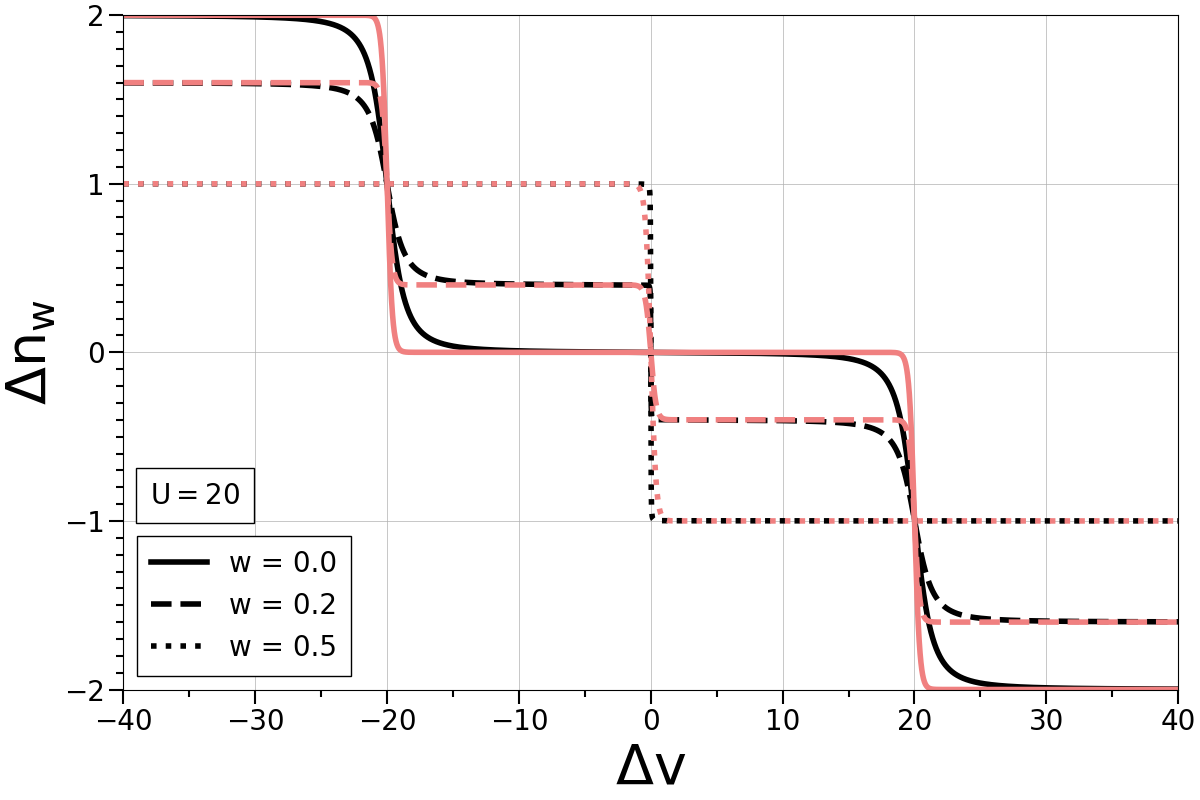}
    \caption{Smooth approximation for the density \ref{eq:SmoothLOn} (light red) and the exact density (black) \ref{coeffs} plotted against $\Delta v$. }
    \label{fig:LO_site_occs}
\end{figure}

Finally we subtract the remaining components to find the correlation energy:
\begin{multline} E\cw(\Delta n\w(x)) \approx U\left(g^{(0)}\w(x) - \frac{x \Delta n\w(x)}{2} - e_{Hx,w}(\Delta n\w(x)) \right)    \\  -T_{s,w}(\Delta n\w(x)) + \frac{g^{(2)}\w(x)}{U}, \label{eq:EcwSC}
\end{multline}
where $e_{Hx,w}(\Delta n\w) = E_{Hx,w}(\Delta n\w)/U$.
Including only lowest order, and inserting the smooth density, Eq.~\ref{eq:SmoothLOn}, yields the plot in figure Fig. \ref{fig:Ecw_smooth}, and the curves in Fig. 5 of the main text. We also plot the the next order contribution to the correlation energy, Eq.~\ref{eq:EcwSC}, in Fig.~\ref{fig:Ecw_smooth_HO}. To avoid divergences from $\alpha$ at $x=\pm1$ we use a smooth approximation to the absolute values that appear in the denominator of Eq.~\ref{eq:gw2}.

\ssec{Correlation Energy on the Adiabatic Connection}
We produce an expression for $E^{\lambda}\cw(\Delta n\w)$ where $\Delta n\w$ is kept fixed for each $\lambda$ along the adiabatic connection. For sufficiently large $U$ we neglect all terms of $\mathcal{O}(1/U^{2})$ and lower. In this limit, 
$\Delta n\w(x) \to \Delta n^{(0)}\w(x)$, and by the adiabatic connection construction we have the requirement that $\Delta v/U \approx \Delta v^{\lambda}/(\lambda U)$, where $\Delta v^{\lambda}$ is the $\lambda$-dependent potential that keeps $\Delta n\w$ fixed along the connection. As a consequence, to leading-order,
\ben \Delta v^{\lambda}(\Delta n\w) \approx \lambda \Delta v^{(0)}(\Delta n\w), \label{eq:dvAC}\een
and thus,
\ben E\w^{\lambda}(\Delta n\w)  \approx \lambda U g^{(0)}\w(\xz) + \frac{g\w^{(2)}(\xz)}{\lambda U}, \label{eq:EwSCAC}\een
where $x^{(0)}(\Delta n\w) = \Delta v^{(0)}(\Delta n\w)/U$ is the inversion of the leading-order density-potential map Eq.~\ref{eq:dvAC}. To produce the correlation energy we subtract from Eq.~\ref{eq:EwSCAC} the external potential energy (along the connection curve), the KS kinetic energy Eq.~\ref{eq:Tsw}, and the HX energy Eq.~\ref{eq:EHXw},
\begin{multline} E^{\lambda}\cw(\Delta n\w) \approx \lambda U\Big(g^{(0)}\w(\xz) - \frac{\xz \Delta n\w}{2} \\ - e_{Hx,w}(\Delta n\w) \Big)    -T_{s,w}(\Delta n\w) + \frac{g^{(2)}\w(\xz)}{\lambda U}, \label{eq:EcwSClam}
\end{multline}
where $\Delta n\w$ remains fixed for all $\lambda$.

\begin{figure}[ht!]
    \centering
    \includegraphics[width=0.48\textwidth]{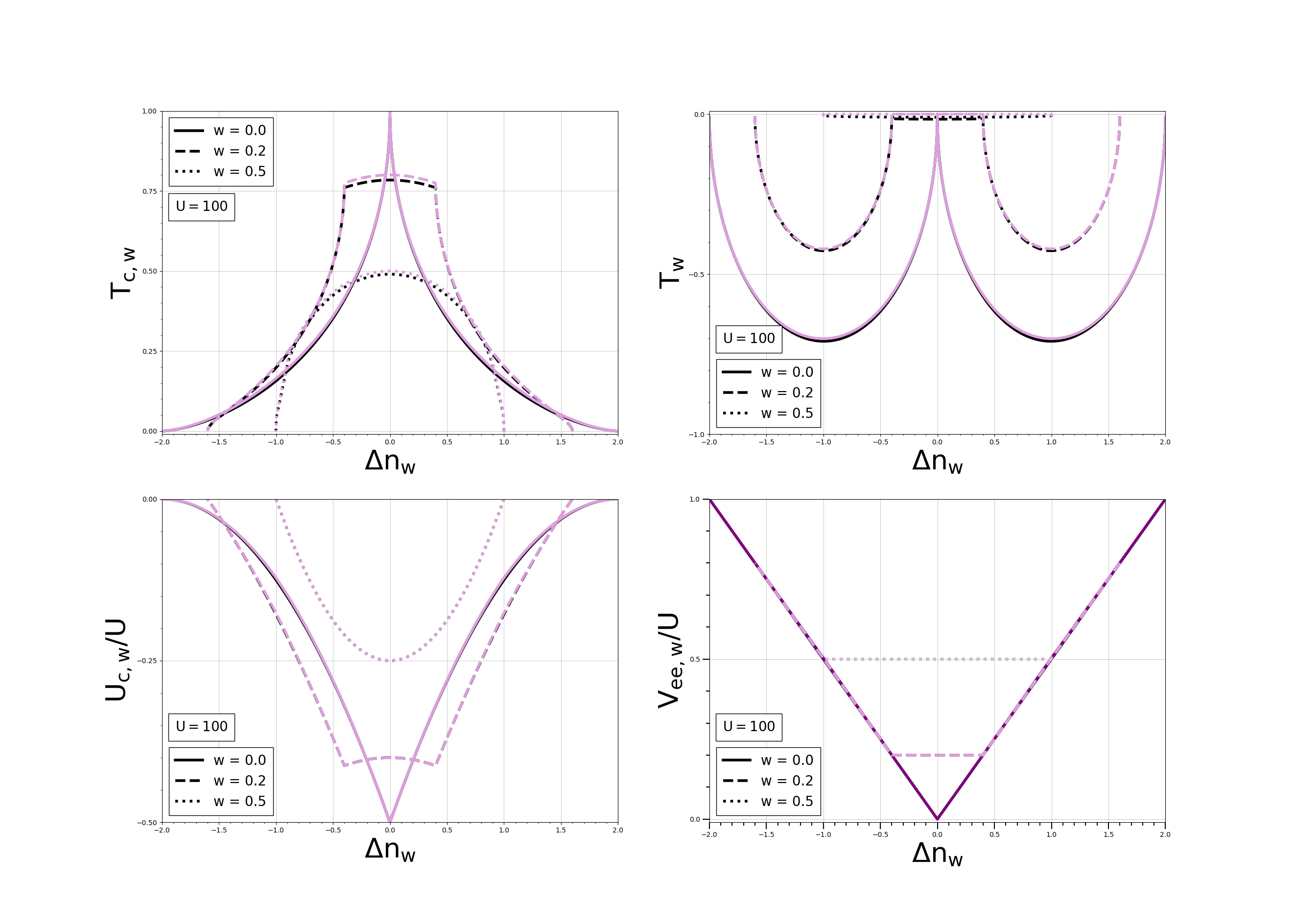}
    \caption{$T_{c,w}$ (\ref{eq:TcSC}), $U_{c,w}$  (\ref{eq:UcSC}), $T_{w}$ (\ref{eq:TSC}), and $V\eew$  (\ref{eq:VeeSC}) for various values of $w$ and $U=100$. The exact values in the strongly correlated limit are represented by the black curves, which are exact as $U\to\infty$.}
    \label{fig:strong-corr1}
\end{figure}

\ssec{Contributions to the Energy}
Equation 20 of the main text yields expressions for the separate kinetic and potential contributions to the correlation energy,
\ben T\cw(\Delta n\w) \approx \frac{2g^{(2)}\w(\xz)}{U} - T_{s,w}(\Delta n\w), \label{eq:TcSC} \een
\begin{multline} U\cw(\Delta n\w) \approx U\Big( g\w^{(0)}(\xz) - \frac{\xz\Delta n\w}{2} \\ - e_{Hx,w}(\Delta n\w) \Big)  - \frac{g^{(2)}\w(\xz)}{U} . \label{eq:UcSC}\end{multline}
From the separate contributions of the correlation energy we deduce that,
\ben T\w(\Delta n\w) \approx \frac{2g^{(2)}\w(\xz)}{U}, \label{eq:TSC}\een
\begin{multline}
 V\eew(\Delta n\w) \approx U\left(g\w^{(0)}(\xz) - \frac{\xz\Delta n\w}{2}\right)  \\ - \frac{g^{(2)}\w(\xz)}{U} . \label{eq:VeeSC} \end{multline}
We plot Eqs.~\ref{eq:TcSC}-\ref{eq:VeeSC} in Fig.~\ref{fig:strong-corr1} with the exact expressions in black and the approximate expressions evaluated with the smooth density in purple. In all plots we use the same smooth approximation to the absolute values in the denominators of $g\w^{(2)}(x)$, Eq.~\ref{eq:gw2}. There are errors in the plots of the correlation kinetic energy, but these vanish as $U\to \infty$.

We compare our result in Eq.~\ref{eq:EcwSC}, for $\vert \Delta n\w \vert \leq 2w$, with the symmetric limit expansion of Deur et al. \cite{deur2018exploring}. To properly compare our approximate correlation energy to the previously reported expansion in the symmetric limit, we produce the weight-dependent constant that vanishes in the limit $U \to \infty$,
\ben 
\frac{E\cw(\Delta n\w)}{U} \approx \frac{\wb}{U} - \frac{1}{2} \left(\wb - \frac{(3 w-1)}{ \wb^2}\frac{\Delta n\w^{2}}{4} \right), \label{eq:EcwDeur}
\een 
which is derived following the procedure in Ref~\cite{deur2018exploring}.
In Fig. 5 of the main text we plot our approximate leading-order correlation energy Eq.~\ref{eq:EcwSC} evaluated with smooth density along with Eq.~\ref{eq:EcwDeur}.

\clearpage
\sec{Ensemble Hartree-Fock Approximation}
We now turn our attention to the ensemble generalization of Hartree-Fock. We define the Hartree-Fock solution of each state (analogous to Eq.~\ref{HubbardWF}) to be
\begin{equation}
    \ket{\Phi_i} = \alpha_i\HF \left(\ket{12} + \ket{21}\right) + \beta_i^{+{\rm HF}} \ket{11} + \beta_i^{-{\rm HF}} \ket{22},
\end{equation}
with coefficients determined to first order in U:
\begin{align}
    \alpha_0\HF &= 2t / c_0\HF \quad\quad &\alpha_1\HF &= -\Delta v_{{\rm eff},w}^2 / 2tc_1\HF \notag \\
    \beta_0^{\pm{\rm HF}} &= 1/2 \pm \Delta v_{{\rm eff},w} / c_0\HF \quad\quad &\beta_1^{\pm{\rm HF}} &= \pm \Delta v_{{\rm eff},w} / c_1\HF \notag \\
    c_0\HF &= 2 \sqrt{4t^2 + \Delta v_{{\rm eff},w}^2} \quad\quad &c_1\HF &= \Delta v_{{\rm eff},w} \sqrt{2 + \Delta v_{{\rm eff},w}^2 / 2t^2} \notag
\end{align}
Here the weight-dependent effective mean-field potential $\Delta v_{{\rm eff},w}$ takes the form \cite{deur2018exploring}
\ben
   \Delta v_{{\rm eff},w} = \Delta v +\frac{(1 - 3w)}{\wb^2} \frac{U \Delta n\w\HF}{2}
   \label{veff}
\een
where $\Delta n\w\HF$ is found from (\ref{coeffs}) with coefficients as above.

\ssec{Densities and Total Energy}

 The self-consistent EHF density is found numerically throughout this work by solving (\ref{veff}).
Plots of the exact/EHF self-consistent site-occupation differences are included below in Fig.~\ref{fig:DnVsDv_HF} for various interaction strengths, $U=0,\,1,\,\&\,5$. Here (and for the remainder of this section) we denote the exact solution using solid curves, and the EHF approximation using dashed curves. Looking at the left panel ($U=0$) of Fig.~\ref{fig:DnVsDv_HF}, the EHF approximation matches the exact
density exactly, as expected in the limit of weak correlation. One can see that the exact/EHF densities (regardless of weight) begin to differ as $U$ is increased, but must always match at the origin (where $\Delta v = \Delta n_w = 0$) and as $\Delta v \to \infty$ (where $|\Delta n_w| = 2\wb$). This behavior would be expected to hold for larger $U$ values, although as noted previously in Ref.~\citenum{deur2018exploring}, there exists a critical interaction strength $U_{\rm crit}$ at which nonphysical behavior is observed for symmetric dimers with bi-ensemble weight $w\geq 1/3$. This critical interaction strength is 
\begin{equation}
    U_{\rm crit} = \frac{\wb}{3w-1}.
\end{equation}
For $w\rightarrow \frac{1}{2}$,  $U_{\rm crit} \rightarrow 1$. At this point the energy expression for dimers with $U>U_{\rm crit}$ have multiple degenerate minima. This explains the deviation from expected behavior for $w=0.4,\,0.5$ in the right panel of Fig.~\ref{fig:DnVsDv_HF}, where both curves approach a finite value as $\Delta v \to 0.$

We also plot the exact/EHF bi-ensemble total energy below, using two interaction strengths ($U=1$ and $U=5$) to examine the EHF approximation in more detail. We depict this quantity as a function of $\Delta v$ below in Fig.~\ref{fig:EnVsDv_HF} and separate each $w$ value in new plots to better illustrate the weight-dependence of $E\w$ and $E\w\HF$.

Analyzing Fig.~\ref{fig:EnVsDv_HF}, one can see that the EHF approximation obeys the variational principle for all viable weight values (where $w \leq 0.5$); as the weight of the first-excited state is increased, both the exact/EHF energy becomes more positive for all $\Delta v$. We note that the EHF approximation always approaches the exact energy as $\Delta v \to \infty$, as shown previously for the ground state.~\cite{Carrascal2015}

\begin{figure}[ht!]
    \centering
    \includegraphics[width=0.48\textwidth]{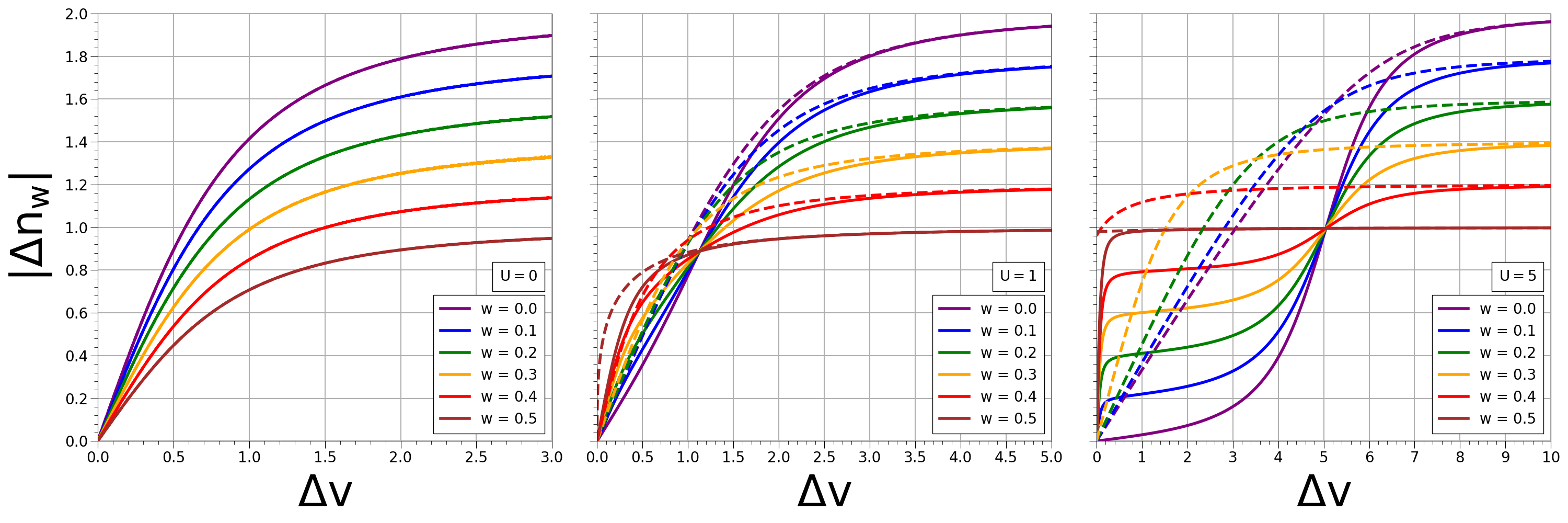}
    \caption{Absolute value of the density of the Hubbard dimer bi-ensemble, plotted for various weight values as a function of $\Delta v$. Here we set $U=0$ (left), $U=1$ (center), and $U=5$ (right). Dashed curves represent the ensemble HF approximation, and the solid curves are exact.}
    \label{fig:DnVsDv_HF}
\end{figure}


\ssec{Correlation Energy}

Below we provide plots of the total weight-dependent correlation energy, as well as its kinetic/potential contributions, in Figs.~\ref{fig:EcVsDv}, \ref{fig:TcVsDv}, and \ref{fig:VcVsDv}. Note that the definition of the EHF correlation energy is
\begin{equation}
    E\HF\cw = E\w[n\w] - E\w\HF[n\w\HF],
\end{equation}
where each energy functional has been minimized by its respective weight-dependent self-consistent density.

Looking at Fig.~\ref{fig:EcVsDv}, it is evident that the behavior of the correlation energy greatly depends upon the value of $w$. We find that the inequality relating the exact/approximate correlation energy holds for all ensembles (i.e. $E\HF\c \geq E\c$ for all weights). We show that a different trend exists for the kinetic/potential correlation components, as the inequalities describing the ground state ($T\HF\c \leq T\c$ and $U\HF\c \leq U\c$) no longer apply for ensembles with $w \neq 0$. 

Note that for each of these quantities, the EHF approximate correlation energy matches the exact solution at $\Delta v = 0$, except for strongly correlated systems with $w \geq 1/3$ (due to the nonphysical behavior in this regime discussed previously).

\begin{figure}[ht!]
    \centering
    \includegraphics[width=0.48\textwidth]{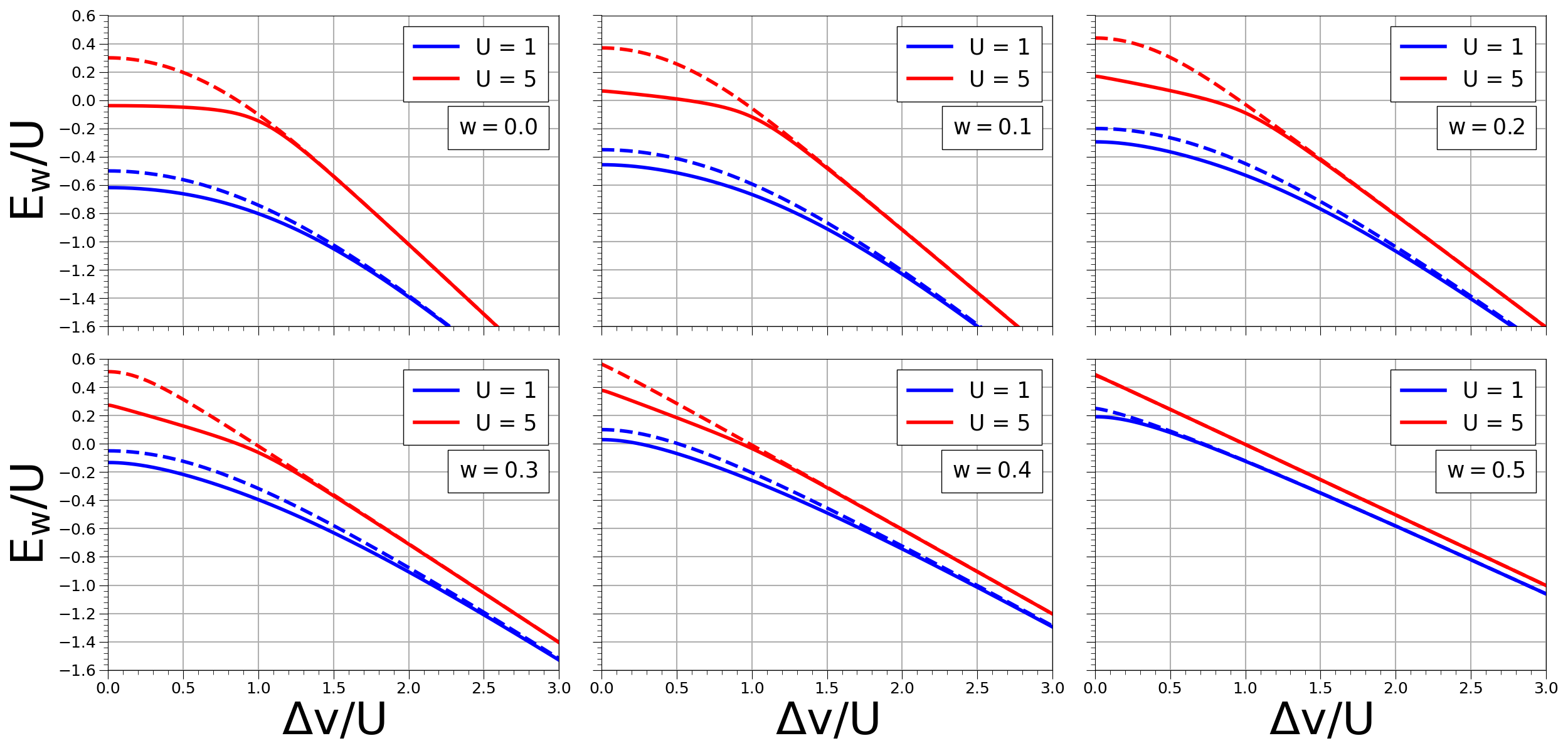}
    \caption{Total energy of the Hubbard dimer bi-ensemble plotted as a function of $\Delta v$ for various $w$ values. Dashed lines are the HF approximation and the solid lines are exact.}
    \label{fig:EnVsDv_HF}
\end{figure}

\begin{figure}[ht!]
    \centering
    \includegraphics[width=0.48\textwidth]{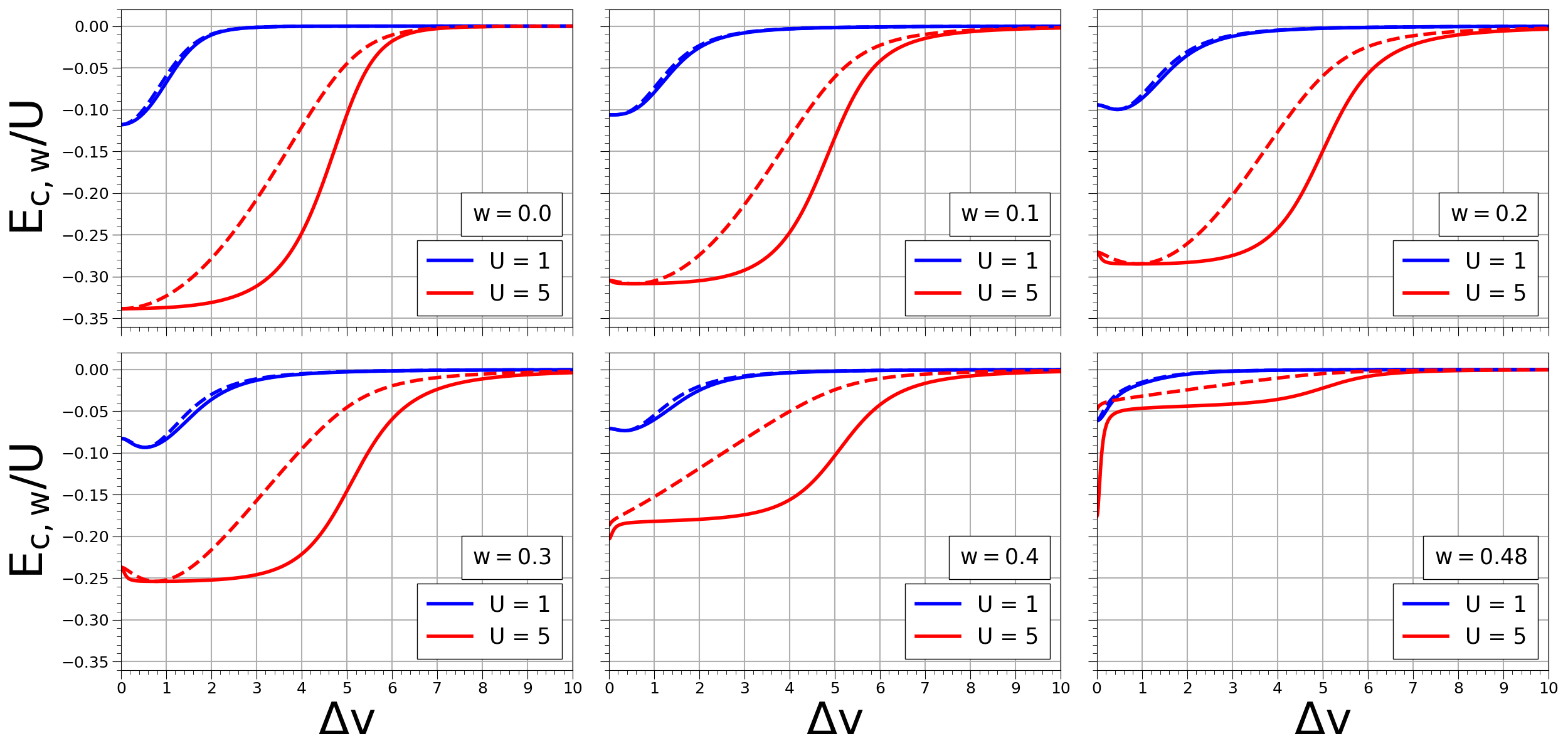}
    \caption{Total correlation energy of the Hubbard dimer bi-ensemble plotted as a function of $\Delta v$ for various $w$ values. Dashed lines are the HF approximation and the solid lines are exact.}
    \label{fig:EcVsDv}
\end{figure}

\begin{figure}[ht!]
    \centering
    \includegraphics[width=0.48\textwidth]{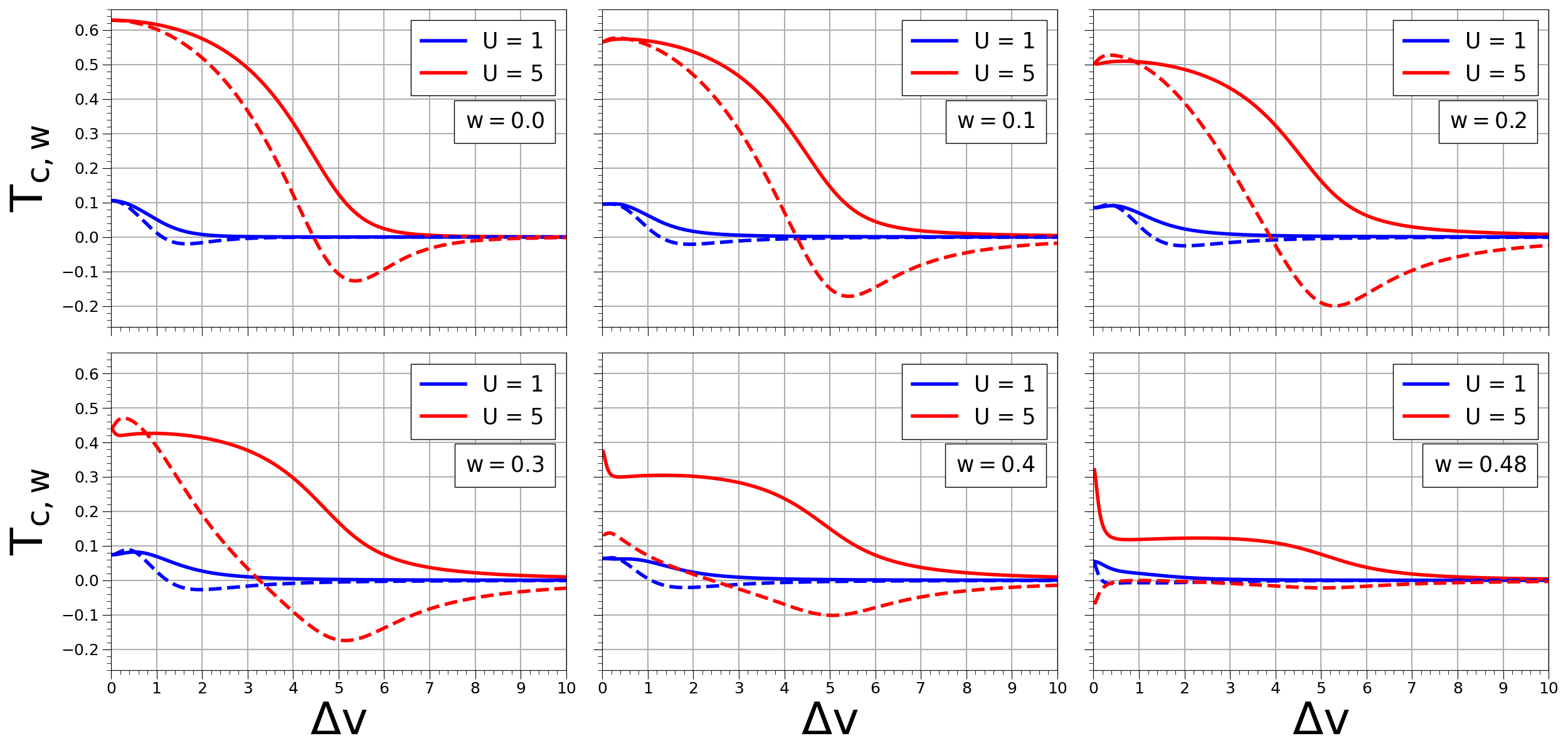}
    \caption{Kinetic correlation energy of the Hubbard dimer bi-ensemble plotted as a function of $\Delta v$ for various $w$ values. Dashed lines are the HF approximation and the solid lines are exact.}
    \label{fig:TcVsDv}
\end{figure}

\begin{figure}[ht!]
    \centering
    \includegraphics[width=0.48\textwidth]{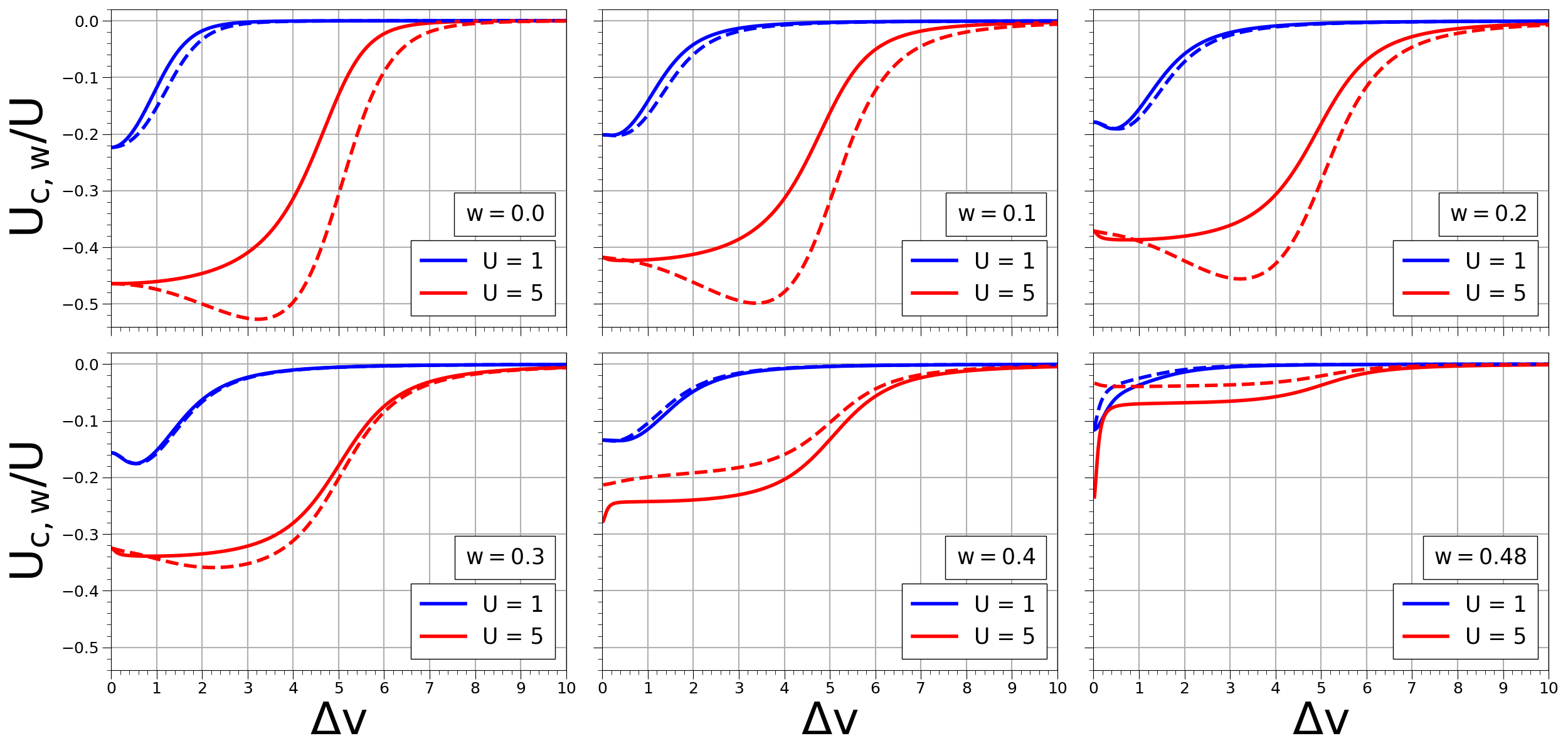}
    \caption{Potential correlation energy of the Hubbard dimer bi-ensemble plotted as a function of $\Delta v$ for various $w$ values. Dashed lines are the HF approximation and the solid lines are exact.}
    \label{fig:VcVsDv}
\end{figure}

\clearpage
\bibliographystyle{apsrev4-2}

\bibliography{main}